\documentclass[prd,aps,a4paper,twocolumn,eqsecnum,nofootinbib,floatfix]{revtex4}  %

\newif\ifusesec
\usesectrue  
   
\usepackage{graphicx} 
\usepackage{amsmath,amsfonts,amssymb}
\usepackage{color}

\newcommand{\eq}{\eqref}
\newcommand{\be}{\begin{equation}}
\newcommand{\ee}{\end{equation}}
\newcommand{\beq}{\begin{equation}}
\newcommand{\eeq}{\end{equation}}
\newcommand{\bea}{\begin{eqnarray}}
\newcommand{\eea}{\end{eqnarray}}
\newcommand{\mb}{\bar m}
\newcommand{\g}{\gamma}
\newcommand{\pinf}{p_{\infty}}
\newcommand{\nn}{\nonumber}
\newcommand{\D}{\partial}
\newcommand{\eps}{\epsilon}
\newcommand{\om}{\omega}

\begin{document}

\title{Gravitational Bremsstrahlung Waveform at the fourth Post-Minkowskian order \\ and the second Post-Newtonian level}

\author{Donato Bini$^{1,2}$, Thibault Damour$^3$, Andrea Geralico$^1$}
  \affiliation{
$^1$Istituto per le Applicazioni del Calcolo ``M. Picone,'' CNR, I-00185 Rome, Italy\\
$^2$INFN, Sezione di Roma Tre, I-00146 Rome, Italy\\
$^3$Institut des Hautes Etudes Scientifiques, 91440 Bures-sur-Yvette, France
}
\date{\today}

\begin{abstract}
Using the Multipolar Post-Minkowskian formalism, we compute the frequency-domain waveform generated by the 
gravitational scattering of two nonspinning bodies at the fourth post-Minkowskian order ($O(G^4)$, or two-loop order),
and at the fractional second Post-Newtonian accuracy ($O(v^4/c^4)$). The waveform is decomposed in
spin-weighted spherical harmonics and the needed radiative multipoles, $U_{\ell m}(\om), V_{\ell m}(\om)$,
are explicitly expressed in terms of a small number of master integrals. The basis of master integrals
contains both (modified) Bessel functions, and solutions of inhomogeneous  Bessel equations with Bessel-function
sources. We show how to express the latter in terms of Meijer G functions. The low-frequency expansion of our results
is checked againg existing classical soft theorems. We also complete our previous results 
on the $O(G^2)$ bremsstrahlung waveform by computing the $O(G^3)$ spectral densities of radiated energy and 
momentum, in the rest frame of one body, at the thirtieth order in velocity.
\end{abstract}

\maketitle

\section{Introduction}

The gravitational-wave (GW) emission from gravitationally interacting binary systems has been the focus of intense
theoretical research over many years because of its relevance to the network of  interferometric GW detetectors.
See \cite{Blanchet:2013haa} for a review.
Recently, a renewed theoretical effort, utilizing advances in quantum scattering amplitudes methods, has been aimed at 
computing the gravitational waveform emitted by the scattering of two massive bodies \cite{Luna:2017dtq,Jakobsen:2021smu,Jakobsen:2021lvp,Mougiakakos:2021ckm,Riva:2022fru,Cristofoli:2021vyo,Adamo:2022qci,Brandhuber:2023hhy,Herderschee:2023fxh,Elkhidir:2023dco,Georgoudis:2023lgf,Gonzo:2023cnv,Aoude:2023dui,Caron-Huot:2023vxl,Georgoudis:2023eke,Georgoudis:2023ozp,Georgoudis:2024pdz,DeAngelis:2023lvf,Bini:2024rsy,Brunello:2024ibk}. 
The current  accuracy on these 
bremsstrahlung waveforms is the one-loop level corresponding to the third post-Minkowskian (3PM) order ($O(G^3)$).
In terms of the classical waveform~\footnote{We use the mostly plus signature,
 $\eta_{\mu \nu}={\rm diag}(-1,+1,+1,+1)$.}
 $h_{\mu \nu} = g_{\mu \nu}-\eta_{\mu \nu}$ we have the PM expansion
\bea 
\label{hmunu}
h_{\mu \nu}(x^\lambda) &=&G h^{\rm lin}_{\mu \nu}+ G^2 h^{\rm 2PM \, or\, tree}_{\mu \nu} + G^3 h^{\rm 3PM\, or\,  one-loop}_{\mu \nu} \nonumber\\
&+&  G^4 h^{\rm 4PM \, or\,  two-loop}_{\mu \nu}  +O(G^5)\,,
\eea
where we indicated the dictionary between the quantum nomenclature (tree, one-loop,...) and the classical PM one.

 Anticipating on the ongoing effort of computing the {\it two-loop} waveform ($G^4$ contribution to $h_{\mu \nu}$,
 denoted $h_{\mu \nu}^{G^4}$), the aim
 of this paper is to present the first five orders, namely $\eta^0+ \eta^1+\eta^2 + \eta^3+\eta^4$ 
 (with $\eta \equiv \frac1c$) in the small-velocity expansion of the classical 4PM (two-loop) waveform, in the frequency
 domain. We recall that each power of $\eta \sim \frac{v}{c}$ is traditionally referred to as being {\it half} a 
 post-Newtonian (PN) order, so that our present $O(\eta^4)$ accuracy corresponds to a fractional 2PN accuracy.
 We hope that the results presented here will provide useful benchmarks when two-loop waveform results
 will be obtained.

Denoting the complex  asymptotic  waveform as
 \bea
 h_c(T_r,\theta,\phi) &=&\lim_{R\to \infty} \eps^{\mu } \eps^{\nu }R\, h_{\mu \nu}\nonumber\\
&=&\lim_{R\to \infty}(R( h_+ -  i h_\times))\,,
 \eea
 where $\eps^\mu = \frac1{\sqrt{2}} (e_\theta^\mu - i e_\phi^\mu)$ is a null polarization vector,
 we present below the value of its $O(G^4)$ frequency-domain contribution at the $\eta^0+ \eta^1+\eta^2 + \eta^3+\eta^4$ accuracy:
 \bea
h^{G^4}_c(\omega,\theta,\phi)&=&\int_{- \infty}^{+\infty}  dT_r e^{i \omega T_r}   h^{G^4}_c(T_r,\theta,\phi)\nonumber\\ 
&=& G^4\left[\eta^0 + \eta^1+ \eta^2 + \eta^3+\eta^4+ O(\eta^5)\right]\,.\nonumber\\
 \eea
 Here, we work in a Bondi-like coordinate system $T_r, R, \theta,\phi$, anchored on the center-of-mass (cm) of
 the binary system. In particular,  $T_r \simeq t- \frac{r}{c} -  \frac{2G {\cal M}}{c^3} \ln \frac{r}{c b_0}$ denotes a retarded time which contains a logarithmic shift proportional to the total mass-energy of the system, $ {\cal M} \equiv \frac{E}{c^2}$, and which depends on the choice of an  arbitrary time scale $ b_0$.
 Denoting the direction of GW emission as ${\bf n}(\theta,\phi) = (\sin \theta \cos \phi, \sin \theta \sin \phi,  \cos \theta)$
 we use as null polarization vector 
 \bea 
\label{eps_def}
{\boldsymbol \eps}=\bar {\bf m}&=& \frac{1}{\sqrt{2}} \left(\D_\theta {\bf n}(\theta,\phi) - \frac{i}{\sin \theta} \D_\phi {\bf n}(\theta,\phi)\right)\,.
\eea

The method we use here to compute $h_c^{G^4}$ is the (PN-matched) Multipolar Post-Minkowskian (MPM) formalism 
 Refs. \cite{Blanchet:1985sp,Blanchet:1986dk,Blanchet:1987wq,Blanchet:1989ki,Damour:1990gj,Damour:1990ji,Blanchet:1992br,Blanchet:1995fr,Poujade:2001ie}. 
This formalism (which is a vast generalization of the classic quadrupole-radiation
 formula of Einstein \cite{Einstein:1918btx}), 
has allowed one to reach a high PN accuracy \cite{Blanchet:2023bwj,Blanchet:2023sbv}
 on the time-domain waveform,
 $ h_c(T_r,\theta,\phi)$, expressed as a PM-expanded, {\it retarded functional}\footnote{This functional comprises various
 linear and nonlinear tail effects, as well as instantaneouslike nonlinear contributions, see \cite{Blanchet:2013haa} for a 
 detailed review, and \cite{Bini:2023fiz} for a streamlined presentation.}
  of some multipole moments of the material source. The latter source multipole moments are essentially defined as a regularized version of the irreducible
 multipole moments of the total (matter + gravitational), conserved stress-energy tensor, $\tau^{\mu\nu}= |g|\, (T^{\mu\nu}_{\rm mat}+ T^{\mu\nu}_{\rm grav})$. 
 
The expression of the time-domain waveform $h_c(T_r,\theta,\phi)$ as a retarded functional
of the source multipoles is not the end of the story. Indeed, one must still: (i) evaluate the source multipole moments
for the considered case of a binary system moving along hyperboliclike trajectories; and (ii) compute the Fourier transform
of the time-domain waveform. As in our previous works \cite{Bini:2023fiz,Bini:2024rsy} we shall use for these
computations the explicit quasi-Keplerian representation of the (PN-expanded) general-relativistic solution for
scattering motions \cite{DD85,Damour:1988mr,Cho:2018upo}.  

For brevity, we shall not repeat here the mini-reviews of the structure of the MPM formalism,
and of the quasi-Keplerian representation, recently presented in \cite{Bini:2023fiz,Bini:2024rsy} 
(see Secs II and IV of \cite{Bini:2023fiz}, and Secs. II and III of \cite{Bini:2024rsy}). We will just
review the notation used for the multipolar decomposition of the waveform  $h_c(T_r,\theta,\phi)$.

We start by recalling that it is technically convenient to factor out of the complex waveform
the  overall factor $ \frac{ 4G}{c^4}  \equiv 4 G \eta^4$ which appears in the classic
(leading PN order) Einstein quadrupole formula:
\be
\eps^i \eps^j h^{\rm LO}_{ij}(t,r,{\bf n}) \approx \frac{4 G}{c^4} \frac{\frac12 \eps^i \eps^j I_{ij}^{(2)}(t-r/c)}{r}\,,
\ee
where $I_{ij}(t) \approx\int d^3 x \frac{T^{00}}{c^2} (x^ix^j- \frac13 {\bf x}^2 \delta^{ij})$ is the
quadrupole moment of the mass-energy distribution, and where the superscript ${}^{(2)}$ denotes a
second time derivative.

When writing $ h_c(T_r,\theta,\phi) = \lim_{R\to \infty} \eps^{\mu } \eps^{\nu }R\, h_{\mu \nu} $
in the factorized form
 \beq
 h_c(T_r,\theta,\phi) \equiv 4 G \eta^4  W(T_r,\theta,\phi)\,,
 \eeq
 the rescaled waveform $ W(T_r,\theta,\phi)$ has, at leading PN order, the {\it Newtonian-level} value 
 $\frac12 \eps^i \eps^j I_{ij}^{(2)}(T_r)$. The MPM formalism obtains $W(T_r,\theta,\phi)$ as a multipolar
 series labeled by the total angular momentum $\ell$, combined with a double series in $G$ and in
 $\eta =\frac1c$. Namely, one has, say after taking the Fourier transform over $T_r$,
 $W(\omega, \theta,\phi) \equiv \int dT_r e^{i \om T_r} W(T_r,\theta,\phi)$,
 \bea \label{MPM}
W_{\rm MPM}(\omega, \theta,\phi)&=& U_2+ \eta (V_2 +U_3) + \eta^2 (V_3+U_4)\nonumber\\ 
&+& \eta^3 (V_4+U_5)+ \eta^4 (V_5+U_6)\nonumber\\
&+& O(\eta^5)\,,
\eea
where we use the standard definitions and normalizations of the various $2^\ell$ multipole contributions,
see, e.g., \cite{Blanchet:2013haa} or, for a recent review, our previous work \cite{Bini:2023fiz}.

Each even-parity, $U_\ell(\om)$, or, odd-parity,  $V_\ell(\om)$, {\it radiative} multipole starts
(for $\om\neq0$) at PM order $G$
and at Newtonian order $\eta^0$. The various factors $\eta^n$ in Eq. \eq{MPM} indicate that, e.g.,
the Newtonian-level  $G^1\eta^0$ term in $U_\ell$ contributes at order  $G^1\eta^{\ell-2}$ to $W$
and  the  $G^1\eta^0$ term in $V_\ell$ contributes at order  $G^1\eta^{\ell-1}$ to $W$.

Here, we are considering the nonstationary part of the waveform,
corresponding to nonzero frequencies $\om$. For instance, at leading PN order, 
this is equivalent, modulo a factor $- i \om$, to considering the Fourier transform of 
$\frac12 \eps^i \eps^j I_{ij}^{(3)}(T_r)$. The latter involves the third derivative of the quadrupole moment,
which vanishes for free motions, and therefore contains a factor $G^1$.
More precisely,
\bea
U_\ell(\om,  \theta,\phi) &=& U_\ell^G(\eta^0+ \eta^2+\eta^4 + \cdots)\nonumber\\ 
&+&  U_\ell^{G^2}(\eta^0+ \eta^2+\eta^3+\eta^4 + \cdots)\nonumber\\
&+& U_\ell^{G^3}(\eta^0+ \eta^2+\eta^3+\eta^4 + \cdots)\nonumber\\
&+&\cdots\nonumber\\
V_\ell(\om,  \theta,\phi) &=& V_\ell^G(\eta^0+ \eta^2+\eta^4 + \cdots) \nonumber\\
&+&  V_\ell^{G^2}(\eta^0+ \eta^2+\eta^3+\eta^4 + \cdots)\nonumber\\
&+& V_\ell^{G^3}(\eta^0+ \eta^2+\eta^3+\eta^4 + \cdots)\nonumber\\
&+&\cdots\,.
\eea
Recent works \cite{Bini:2023fiz,Bini:2024rsy} have displayed the building blocks 
$U^G_\ell(\om,  \theta,\phi)$, $U^{G^2}_\ell(\om,  \theta,\phi)$ and 
$V^G_\ell(\om,  \theta,\phi)$, $V^{G^2}_\ell(\om,  \theta,\phi)$ needed to reach the
overall accuracy\footnote{When computing the accuracy of the waveform $h_c(\om,\theta,\phi)$,
we count the overall power of $G$, but we discount the overall factor $\eta^4$ to focus
on the fractional PN accuracy.} $G^3 \eta^5$ for the even-in-$\phi $ part of the waveform  $h_c(\om,\theta,\phi) \equiv 4 G \eta^4  W(\om,\theta,\phi)$. [The odd-in-$\phi$ part of the waveform was given with the reduced accuracy $G^3 \eta^3$.]
This corresponds to the accuracy  $U_2 \sim (G+G^2)(\eta^0 + \eta^2 +\eta^3+\eta^4+\eta^5)$ for $\ell=2^+$,
the accuracy   $V_2  \sim U_3 \sim (G+G^2)(\eta^0 + \eta^2 +\eta^3)$ for $\ell=2^-$ and $\ell=3^+$,
the accuracy   $V_3  \sim U_4 \sim (G+G^2)(\eta^0 + \eta^2 +\eta^3)$ for $\ell=3^-$ and $\ell=4^+$,
the accuracy   $V_4  \sim U_5 \sim (G+G^2)(\eta^0 + \eta^2)$ for $\ell=4^-$ and $\ell=5^+$ (not given in our previous work),
and the accuracy   $V_5  \sim U_6 \sim (G+G^2)(\eta^0)$ for $\ell=4^-$ and $\ell=5^+$.

Here, our aim is to reach the overall accuracy  $G^4 \eta^4$ (4PM and 2PN) on the $c^4$-rescaled  waveform
$\eta^{-4} h_c(\om,\theta,\phi) \equiv 4 G   W(\om,\theta,\phi)$. This will be obtained by having the
following accuracies on the $U^{G^3}_\ell(\om,  \theta,\phi)$ and $V^{G^3}_\ell(\om,  \theta,\phi)$:
\bea
U_2^{G^3} &\sim& G^3(\eta^0 + \eta^2 +\eta^3+\eta^4)\,,\nonumber\\
V_2^{G^3} &\sim& U_3^{G^3} \sim G^3(\eta^0 + \eta^2 +\eta^3)\,,\nonumber\\
V_3^{G^3} &\sim&  U_4^{G^3}\sim G^3(\eta^0 + \eta^2 )\,,\nonumber\\  
V_4^{G^3} &\sim& U_5^{G^3} \sim V_5^{G^3} \sim U_6^{G^3} \sim G^3\eta^0\,. 
\eea
In the following, we focus on the structure of, and result for, the even-parity radiative quadrupole $U_2$,
which requires the highest fractional PN accuracy.

\section{Structure of the 4PM quadrupolar waveform at 2PN fractional accuracy}

The even-parity, quadrupolar contribution $U_2$ to the rescaled waveform 
$W(T_r,\theta,\phi) =\frac{c^4}{4G} h_c(T_r,\theta,\phi)$ reads
\bea
U_2(T_r,\theta,\phi) &=& \frac{1}{2} \mb^{ij}   U_{i j}\,,
\eea
where the radiative quadrupole (measured at future null infinity) is given,
in the MPM formalism, by a retarded functional of the form
\bea \label{Uij}
U_{ij}(T_r) &=& I^{(2)}_{ij} (T_r)\\
&+&
{2G {\mathcal M} \over c^3} \int_{-\infty}^{T_r} d T_r' \left[ \ln
\left({T_r- T_r' \over 2 b_0}\right)+{11 \over12} \right] I^{(4)}_{ij} (T_r')\nonumber\\ 
&+& O\left(\frac{G^2}{c^5}\right)\,.
\eea
Here, the (time-dependent part of the) source multipole $ I_{ij}(t)$ has to be evaluated at the $(G+G^2+G^3)(\eta^0+\eta^2 +\eta^4)$ accuracy (i.e. the 2PN accuracy) in the first term on the right-hand side (rhs), while only its Newtonian-level ($O(\eta^0)$),
but 2PM-accurate ($O(G+G^2)$), value is needed in the second, tail, term on the rhs. We recall that 
${\mathcal M}= \frac{E_{\rm cm}}{c^2}$, and that $b_0$ denotes the arbitrary time scale entering the MPM
formalism (both in the definition of the log contribution to the retarded time, and in the partie-finie regularization
entering the nonlinear MPM iteration scheme). For illustration, the beginning of the 
PN expansion of the (time-domain) MPM source quadrupole moment reads \cite{Blanchet:1989ki} 
\bea \label{Iij1PN}
I_{ij}(t)&=& \int d^3 x  \sigma(t, {\bf x}) {\hat x}^{i j} \nonumber\\
&+& \frac1{14 c^2 } \frac{d^2}{dt^2}\int  d^3 x  \sigma(t, {\bf x}) {\bf x}^2 {\hat x}^{i j}\nonumber\\
 & -& \frac{20}{21 c^2}  \frac{d}{dt} \int  d^3 x  \sigma_k(t, {\bf x}) {\hat x}^{i j k}\nonumber\\
& +& O\left(\frac1{c^4}\right)\,.
\eea
Here, $\sigma=\frac{T^{00}_{\rm mat}+ T^{kk}_{\rm mat}}{c^2}$, $\sigma_k=\frac{T^{0k}_{\rm mat}}{c}$, $ {\hat x}^{i j \cdots}$ denotes the symmetric-trace-free (STF) projection of $x^i x^j \cdots$, and the coordinates $x^i$ refer to a source-rooted harmonic coordinate system.
As in Refs. \cite{Bini:2023fiz,Bini:2024rsy}, the Cartesian components of all multipoles, and the associated emission
angles $\theta, \phi$, are defined with respect to a cm-frame spatial basis $e_x, e_y, e_z$ where $e_x, e_y$
are anchored on the classical averaged cm-frame momenta. More precisely, 
the time axis ${\bar e}_0$ of the cm frame is
\be
\label{eq:ebar0}
{\bar e}_0^\mu= \frac{{\bar p}_1^\mu+{\bar p}_2^\mu}{|{\bar  p}_1^\mu+{\bar p}_2^\mu|}\,.
\ee
The vector $e_y$ lies in the spatial direction of ${\bar p}_1$ (i.e.  the bisector between the incoming and the outgoing spatial momentum of the first particle in the cm frame),
\be
\label{eq:p1p2}
{\bar p}_1 = {E}_1  {\bar e}_0 +  {\bar P}_{\rm c.m.} e_y\,,\qquad {\bar p}_2 = {E}_2  {\bar e}_0 -  {\bar P}_{\rm c.m.} e_y\,,
\ee
where $E_a = \sqrt{m_a^2+P_{\text{c.m.}}^2}$, and $P_{\text{c.m.}}$ is the magnitude of the spatial part of the incoming momentum, while ${\bar P}_{\text{c.m.}}={P}_{\rm c.m.}  \cos \frac12 \chi$.  
The axis vector $e_x$ is in the plane of motion and orthogonal to $e_y$ (and oriented from particle 2 towards
particle 1). It is also the direction of the eikonal vectorial impact parameter, $b_{\text{eik}} e_x$. 
Here $b_{\text{eik}}$ is the eikonal-type impact parameter, linked to the usual, cm, incoming-momenta impact parameter $b$ by
$b_{\text{eik}} = \frac{b }{\cos\frac{1}{2}\chi}$.
The last spatial axis vector $e_z$ is orthogonal to the plane of motion (and such that $e_x$, $e_y$, $e_z$
is positively oriented).
All vectors and tensors are decomposed in the frame  ${\bar e}_0$, $e_x$, $e_y$, $e_z$ and
the angles $\theta,\phi$ are accordingly defined, so that
\bea
\label{frame_exey}
{\bf n}(\theta,\phi) &=& \sin\theta \cos\phi \, e_x +\sin\theta\sin\phi \, e_y +\cos\theta \,e_z \,, \nn\\
k &=& \om \left( {\bar e}_0 + {\bf n}(\theta,\phi) \right), \nn \\
\mb &=& \frac{1}{\sqrt{2}} \left[\D_\theta {\bf n}(\theta,\phi)-  \frac{i}{\sin \theta} \D_\phi {\bf n}(\theta,\phi)\right].
\eea
Here ${\bf n}$ is the spatial unit vector that characterizes the direction of the gravitational wave vector $ k =(\om, {\bf k})$. 

The explicit computation (at the accuracy needed here) of the Fourier-transform of the first contribution to $U_{ij}$ 
then proceeds as follows: (i) one inserts the $T^{\mu\nu}_{\rm mat}$ describing two point masses at 2PN accuracy
(see \cite{Blanchet:1989cri,Blanchet:1989cu});
(ii) one inserts the explicit 2PN-accurate quasi-Keplerian representation of the relative hyperbolic motion, 
$x^i(t)=x_1^i(t)- x_2^i(t)$, namely
\beq
{\bf x}(t)= r(t) (\cos \varphi(t) e_x + \sin \varphi(t) e_y)\,,
\eeq
where the polar coordinates of the relative motion are obtained as functions of time by eliminating the
(hyperbolictype) ``eccentric anomaly" variable $v$ between equations of the type
\bea \label{QK}
{\bar n}\, t &=& e_t \sinh(v) - v + O(\eta^4)\,, \nn\\
r &=& {\bar a}_r \left( e_r \cosh(v) -1) \right) + O(\eta^4)\,, \nn \\
\varphi &=& 2 K \arctan\left(\sqrt{\frac{e_\varphi+1}{e_\varphi-1}}\tanh\left(\frac{v}{2}\right)\right)   + O(\eta^4)\,.\nonumber\\
\eea
Here, the quasi-Keplerian quantities ${\bar n}, e_t, e_r, e_\varphi, K$ are (PN-expanded) functions of the
c.m. energy, $E$, and angular momentum, $J$, of the binary system (see Refs. \cite{DD85,Damour:1988mr,Cho:2018upo}).
[$E$ and $J$ are then reexpressed in terms of the incoming Lorentz factor $\g$ and the (incoming-momenta) impact
parameter $b$.]
The quasi-Keplerian representation Eq. \eqref{QK} incorporates (in the conservative case) a time symmetry around
$t=0$, corresponding to the closest approach between the two bodies. The asymptotic logarithmic drift of the two
worldlines is embodied in the $v$ parametrization involving hyperbolic functions.
  
  Finally, the Fourier transform is conveniently computed by considering $ I^{(3)}_{ij}(\om) = - i \om  I^{(2)}_{ij}(\om) $
  and by taking the {\it large eccentricity} expansion of  the Fourier integrals generated by the quasi-Keplerian 
  representation, which have the form
\be
\int dv e^{i \frac{\om}{{\bar n}} \left(  e_t \sinh(v) - v  \right)} \left( F_0(v) + \eta^2  F_2(v) + \cdots \right)\,.
\ee
The eccentricity $e_t$ (when expressed in terms of the cm energy $E$ and the  cm angular momentum $J$)
is of the form 
\bea \label{et}
 e_t(E,J,G) &\sim& \sqrt{ 1 + \frac{ \pinf^2  J^2}{(G  m_1 m_2 )^2}} \left( 1 \right.\nonumber\\
&+&\left. O\left(\frac{\pinf^2}{c^2}\right) + O(G^2)\right)\,,
\eea
where we parametrize the total cm energy by the  variable $\pinf \equiv c \sqrt{\g^2-1}$ (having the
dimension of a velocity), with $\g$ the relative Lorentz factor between the incoming particles. In view of the expression \eq{et},
the large-eccentricity expansion corresponds to a PM expansion in powers of $G$.

In previous (one-loop-level) works \cite{Bini:2023fiz,Bini:2024rsy}, we only had to consider the first two terms
in the large-eccentricity expansion of the multipole moments. The novelty of the present study is to take
into account the third term, $O(\frac1{e_t^3}) = O(G^3)$, in the large-eccentricity expansion of the
frequency-domain multipoles, as given by the quasi-Keplerian representation \eqref{QK}.
As discussed next, the Fourier integrals appearing at order  $O(\frac1{e_t^3}) = O(G^3)$ introduce
new transcendental functions which go beyond the ones introduced at orders $O(G+G^2)$.
[At orders $O(G+G^2)$ the frequency-domain waveform $W(\om)$ was fully expressible in terms of the two
Bessel K functions $K_0( \frac{\om \, b}{ \pinf})$ and $K_1( \frac{\om \, b}{ \pinf})$,
together with $\exp(-\frac{\om \, b}{ \pinf})$. See Appendix \ref{appD} for
the order $G^1$ and Refs. \cite{Bini:2024rsy,Bini:2023fiz} for the order $G^2$.]

Finally, the frequency-domain radiative quadrupole moment of a scattering binary is obtained as a double expansion
in $G$ and $\eta$ of the form
\bea
U_2(\om, \theta,\phi) &=&U_2^{G^1}(\om, \theta,\phi) + U_2^{G^2}(\om, \theta,\phi)\nonumber\\
&+& U_2^{G^3}(\om, \theta,\phi)+ O(G^4)\,,
\eea
where each PM contribution $O(G^n)$, for $n=1,2,3$, 
has a small-velocity (PN) expansion of the form (at our present accuracy)
\bea
U_2^{G^n}(\om, \theta,\phi)&=& U_2^{G^n \eta^0}+  U_2^{G^n \eta^2}+ U_2^{G^n \eta^3}\nonumber\\
&+& U_2^{G^n \eta^4}+ O(G^n \eta^5)\,.
\eea
The small-velocity expansion is taken at a fixed value of the following dimensionless frequency parameter
\be
u \equiv \frac{\om \, b}{ \pinf}\,.
\ee
Here, $b = \frac{J}{P_{\rm cm}}$ is the  impact parameter.
 
When $U_2(\om)$ (with dimensions $[M V^2 T]$) is expressed in terms of the dimensionless variables $u, \theta, \phi$
the velocity expansions of its successive PM contributions have the structure (where we factor out an overall symmetric
mass ratio $\nu \equiv \frac{m_1 m_2}{M^2}$, with $M \equiv m_1+m_2$)
\bea
U_2^{G^1}(u, \theta,\phi) &=&  \nu  \frac{GM^2}{p_\infty }\left[1+\frac{p_\infty^2}{c^2}+\frac{p_\infty^4}{c^4}  +O\left(\frac{p_\infty^6}{c^6}\right) \right]\nonumber\\
U_2^{G^2}(u, \theta,\phi) &=&  \nu  \frac{GM^2}{p_\infty } \left(\frac{GM}{bp_\infty^2}\right)\left[1+\frac{p_\infty^2}{c^2}+\frac{p_\infty^3}{c^3}\right.\nonumber\\  &+&\left.\frac{p_\infty^4}{c^4}+O\left(\frac{p_\infty^5}{c^5}\right)  \right]\nonumber\\
U_2^{G^3}(u, \theta,\phi) &=&  \nu  \frac{GM^2}{p_\infty }\left(\frac{GM}{bp_\infty^2}\right)^2\left[1+\frac{p_\infty^2}{c^2}+\frac{p_\infty^3}{c^3}\right.\nonumber\\  &+&\left. \frac{p_\infty^4}{c^4}+O\left(\frac{p_\infty^5}{c^5}\right)  \right]\,.
\eea
Here, the dimensionless factor $\left(\frac{GM}{bp_\infty^2}\right)$ keying the PM expansion
 is of order $G$ but is of Newtonian order $\eta^0$ (remember
that  $\pinf$ is a velocity). This factor is of order of the scattering angle $\chi$.

The explicit values of $U_2^{G^1}(u, \theta,\phi)$ and $U_2^{G^2}(u, \theta,\phi)$ have been
given in Refs. \cite{Bini:2023fiz,Bini:2024rsy} (and their ancillary files) to a rather high PN accuracy. 
Here, we consider only the 3PM contribution $U_2^{G^3}(u, \theta,\phi)$ to $W(u, \theta,\phi)$ (corresponding
to the 4PM waveform) at the fractional PN accuracy indicated above, i.e. $1+\frac{p_\infty^2}{c^2}+\frac{p_\infty^3}{c^3}+\frac{p_\infty^4}{c^4}  +O(\frac{p_\infty^5}{c^5})$. The explicit values of the complementary higher multipolar contributions
$V_2,U_3$ (at 1.5PN accuracy),  $V_3,U_4$ (at 1PN accuracy), and $V_4, U_5, V_5, U_6 $ (at the leading Newtonian order) are given
in the ancillary files  of the arxiv submission  of this work.

In order to clarify  the $u$ dependence of the multipolar waveforms, it is convenient to factor out the angular
dependence which is simply given by appropriate spin-weighted spherical harmonics (SWSH). For example,
for $\ell=2^+$, we have
\bea
U_2(u,\theta,\phi)&=& U_{22}(u) Y_{{\bar 2}; 22}(\theta,\phi) + U_{2 0}(u)  Y_{{\bar 2}; 20}(\theta,\phi)\nonumber\\ 
&+& U_{2 \bar{2}} (u) Y_{{\bar 2}; 2{\bar 2}}(\theta,\phi)\,,
\eea 
where, as stated above, each term $U_{2m}(u)$ (for $m=2,0,-2)$ has both a $G$-expansion and an $\eta$-expansion and we used the notation
$Y_{{\bar s}=-s; \ell m}(\theta,\phi)$ for the SWSH.

\section{The 1.5PN, tail contribution to the 4PM waveform}

It is convenient to discuss first the $\frac{p_\infty^3}{c^3}$ (1.5PN) contribution to  $U_2^{G^3}(u, \theta,\phi)$.
This contribution only comes from the  nonlocal-in-time (tail) term in Eq. \eq{Uij} because the PN expansion 
of the local-in-time  term $I^{(2)}_{ij}(t)$ reads $\sim \eta^0+ \eta^2 + \eta^4+\eta^5$. In Fourier-space
the nonlocal tail kernel diagonalizes and becomes a factor (involving Euler's constant $\g_E$ and the sign of the frequency)
\bea \label{ftail}
f^{\rm tail}_{U_2}(\om)&=&\frac{2 G {\mathcal M} |\om|}{c^3} \left[\frac{\pi}2\right.    \nonumber\\
&+&\left.  i \,{\rm sign}(\om)\left(\ln(2 |\om| b_0)+ \g_E  -\frac{11}{12}\right)\right].
\eea
 
The tail factor \eq{ftail} involves the time scale $b_0$ introduced in the definition of the retarded time.

As this factor is $O(G)$, the tail contribution to the $O(G^3)$ quadrupole $U_2$ is obtained by multiplying 
 $U_2^{G^2 \eta^0}$ by $f^{\rm tail}_{U_2}(\om)$. On the other hand, we found in Ref. \cite{Bini:2023fiz} that $U_2^{G^2 \eta^0}$ differed
 from the leading-order term $U_2^{G^1 \eta^0}$ only by a factor 
 \be \label{fG}
 f^{\frac{G^2 \eta^0}{G^1 \eta^0}}(\om)= \frac{\pi}{2} \left(\frac{GM}{bp_\infty^2}\right) \, u \,. 
 \ee
 Finally, the 1.5PN, tail contribution to $U_2^{G^3}(u, \theta,\phi)$ is simply
 \be
 U_2^{G^3 \eta^3}(u, \theta,\phi)= f^{\rm tail}_{U_2}(\om) f^{\frac{G^2 \eta^0}{G^1 \eta^0}}(\om)  U_2^{G^1 \eta^0}(u, \theta,\phi)\,.
 \ee
 Here the leading-order term $U_2^{G^1 \eta^0}$ (written here for simplicity in the equatorial plane, $\theta=\frac{\pi}{2}$)  is
 \bea
U_{2}^{G^1\eta^0}(u, \frac{\pi}{2},\phi)
&=&-\frac{GM^2\nu}{2p_\infty}\left[\left(2iu\sin(2\phi)\right.\right.\nonumber\\
&+&\left.\cos(2\phi)+1\right)K_0(u) \nonumber\\
&+&\left. \left(2i\sin(2\phi)+2u\cos(2\phi)\right)K_1(u)\right]\,.\nonumber\\
\eea
See the ancillary file of Ref. \cite{Bini:2023fiz}.

for the corresponding result outside of the equatorial plane, given
in terms of SWSHs.

Similarly to the tail contribution to $U_2$ shown above, working at the 2PN level of accuracy we have to include 
the tail contributions to $V_2$ and $U_3$.
We find
 \bea
V_2^{G^3 \eta^3}(u, \theta,\phi)&=& f^{\rm tail}_{V_2}(\om) f^{\frac{G^2 \eta^0}{G^1 \eta^0}}(\om)  V_2^{G^1 \eta^0}(u, \theta,\phi)\,,\nonumber\\
U_3^{G^3 \eta^3}(u, \theta,\phi)&=& f^{\rm tail}_{U_3}(\om) f^{\frac{G^2 \eta^0}{G^1 \eta^0}}(\om)  U_3^{G^1 \eta^0}(u, \theta,\phi)\,,\nonumber\\
 \eea
where 
\bea
f^{\rm tail}_{V_2}(\om) &=&  \frac{2 G {\mathcal M} |\om|}{c^3} \left[\frac{\pi}2\right.    \nonumber\\
&+&\left. i \,{\rm sign}(\om)\left(\ln(2 |\om| b_0)\g_E  -\frac{7}{6}\right)\right] 
\,,\nonumber\\
f^{\rm tail}_{U_3}(\om) &=&  \frac{2 G {\mathcal M} |\om|}{c^3} \left[\frac{\pi}2\right.    \nonumber\\
&+&\left.  i \,{\rm sign}(\om)\left(\ln(2 |\om| b_0)\g_E  -\frac{97}{60}\right)\right]
\,.
\eea

\section{Master integrals for the Newtonian, 1PN and 2PN contributions to the 4PM waveform}

We now come to the instantaneous contributions to the waveform. We recall that at order $G^1$ the
SWSH components of the quadrupolar waveform $U_2(u)$ involve (at any velocity order) 
linear combinations of the modified Bessel functions $K_0(u)$
and $K_1(u)$ with coefficients given by polynomials in $u$.  
By contrast, at order $G^2$ there are two types of contributions: some involve linear combinations
of $K_0(u)$ and $K_1(u)$, while others (which start at order  $G^2 \eta^2$) involve the
new transcendental function $\frac{e^{-u}}{u}$, multiplied by a polynomial in $u$.

The origin of these results is the following. First, we recall that the news function $\dot h_c \sim \sum(\dot U_\ell+\dot V_\ell )$ vanishes when evaluated along free motions. It is therefore expressible (when viewed
from a PM perspective) as the product of $G$ with some integral expression involving
the PM expansion of the relative motion (in the cm frame). The latter cm-frame PM expansion is a direct
reflection of the covariant PM expansion of two worldlines: $z_A^\mu= b_A^\mu+ \tau_A u_A^\mu + G z_A^{\mu {\rm 1PM}}+ O(G^2)$.
The $O(G^1)$ value of $W(\om)$ only depends on the insertion in the latter expression of the free-motion solution 
$z_A^\mu= b_A^\mu+ \tau_A u_A^\mu +O(G)$. This is easily seen  to generate (in the cm frame, and at any velocity order)
 Fourier integrals of the generic form ($n$ denoting a positive integer)
\be \label{I1}
I_1(n) =\int dT \frac{e^{iu T}}{(1+T^2)^{n + \frac12}}\,.
\ee
By contrast, the  $O(G^2)$ value of $W(\om)$ depends on the insertion of the $ G z_A^{\mu {\rm 1PM}}$ 
correction to incoming free motions (projected on the cm-frame relative motion), as well as on $O(\frac{GM}{r})$ fractional corrections entering the source multipole
moments. The 1PM correction to the worldline contains three types of terms (see, e.g., \cite{Bini:2022wrq}): 
terms rational in $\tau_A$, algebraic terms involving $D(\tau_A)= \sqrt{b^2+ \pinf^2 \tau_A^2}$, and
transcendental terms involving $\ln (\pinf \tau_A + D(\tau_A) )$. The rational terms yield contributions
of the type $I_1(n)$. The algebraic terms involving $D(\tau_A)= \sqrt{b^2+ \pinf^2 \tau_A^2}$, together
with the  $\propto \frac{GM}{r}$ fractional corrections to the multipole moments (which, in the PN
expansion, start at order  $G^2 \eta^2$)   add  an extra factor $\frac1{(1+T^2)^{ \frac12}}$ in the
first Fourier integrals above, leading to  contributions of the generic form
\be \label{I2}
I_2(n)=\int dT \frac{e^{iu T}}{(1+T^2)^{n + 1}}\,.
\ee
Finally, the transcendental terms yield contributions of the new type
\be \label{I}
I_3(n)=\int dT \frac{e^{iu T}}{(1+T^2)^{n + \frac12}} \, {\rm arcsinh} T\,.
\ee
The first type of integral, $I_1(n)$, is expressible in terms of $K_n(u)$ functions. The second type, $I_2(n)$, 
which can be easily computed as a Cauchy integral, is
expressible in terms of $e^{-u}$. The third type, $I_3(n)$, is expressible in terms of the
{\it first} derivative of a Bessel function $K_\nu(u)$ with respect to the order $\nu$, when taking the limit $\nu \to n$
after differentiation. Then, known identities concerning  $\D_\nu K_\nu(u)$ at integer values of $\nu$ \cite{NIST,Brychkov:2016},
allow  one to re-express $I_3(n)$ in terms of ordinary $K_n(u)$ functions. 
See also Appendix \ref{eval_int} for a direct proof.

When going to the two-loop level ($G^3$  in $W$) one gets more complicated Fourier integrals
defining higher transcendental functions of $u$. Here, we discuss the integrals that appeared at the
limited PN accuracy of our present study. [All the integrals discussed here correspond to some diagrams
of exchanged gravitons between the two particle worldlines. In our approach the exchange of gravitons is encoded
in the quasi-Keplerian solution, and it is the $G$ expansion of the corresponding multipole moments which generate
the corresponding integrals.]  In order to clarify the structure of the resulting waveform, and the possibility
of expressing the waveform in terms of a finite basis of  {\it master integrals}, we took a systematic approach to,
and a more systematic notation for, the appearing integrals.
Namely, let us denote the simpler integrals appearing in Eqs. \eq{I1} and \eq{I2} as
\be 
\label{Q}
Q_{\alpha}(u) \equiv \int dT \frac{e^{iu T}}{(1+T^2)^{\alpha}} \,,
\ee
for a generic index $\alpha$ (possibly complex, if analytic continuations in $\alpha$ helps its discussion).
We generally assume here that the argument $u$ is (strictly) positive.

For generic values of $\alpha$ the special functions $Q_{\alpha}(u)$ defined by Eq. \eq{Q} can be expressed in
terms of modified Bessel $K$ functions, namely
\bea
\label{Q_integ}
Q_{\alpha}(u)
&=& \frac{ 2^{\frac{3}{2}-\alpha}\sqrt{\pi}  u^{-\frac{1}{2}+\alpha} }{\Gamma(\alpha)}
  {\rm BesselK}\left(-\frac{1}{2}+\alpha, u\right)\,. \qquad
\eea
[We recall that Bessel functions of half-integer orders are expressible in terms of exponential functions.]
The integrals appearing in the $G^3$ level of the rescaled waveform $W$ (at our considered PN accuracy)
are all directly expressed, without using any integration by parts (IBP), either in terms of integrals
of the type \eq{Q}, or of the following more complicated integrals:
\bea
\label{all_integrals}
Q_{\alpha}^{\rm at}(u)&=&  \int dT \frac{e^{iu T}}{(1+T^2)^{\alpha}}{\rm arctan}\left(T\right)
\,,\nonumber\\
Q_{\alpha}^{\rm as}(u)&=& \int dT \frac{e^{iu T}}{(1+T^2)^{\alpha}}{\rm arcsinh}(T)
\,,\nonumber\\
Q_{\alpha}^{{\rm as}2}(u)&=& \int dT \frac{e^{iu T}}{(1+T^2)^{\alpha}}{\rm arcsinh}^2(T)\,.
\eea
Here, the index $\alpha$ takes (for the relevant integrals)  half-integer values except for the second integral ($Q_{\alpha}^{\rm as}$), where it takes  integer values.

Note that all those integrals are of the generic form
\beq
I_\alpha^f(u) \equiv \int dT \frac{e^{iu T}}{(1+T^2)^{\alpha}}f(T)\,,
\eeq
with some function $f(T)$. In addition, the functions $f(T)$ entering the first two types of integrals
in \eq{all_integrals} have the special property that the derivative of the extra factor $f(T)$ satisfies
\beq  
\label{f'}
f'(T)=\frac{1}{(1+T^2)^\beta}\,,
\eeq
with $\beta=1$ for the arctan integral and $\beta=1/2$ for the arcsinh integral.

One can derive some simple {\it ladder relations} linking various generic $I_\alpha^f(u)$ integrals.
Introducing the first-order differential operator
\beq
\widehat T =\frac{1}{i}\frac{d}{du}\,,
\eeq
it is easy to see that\footnote{Here and below we assume that the real part of $\alpha$ is large enough
for all the considered integrals to be convergent. Limiting cases must be treated separately (e.g., using
analytic continuation in $\alpha$).}
\beq
\label{prima_rel}
(\widehat T^2+1)I_\alpha^f(u)=I_{\alpha-1}^f(u)\,.
\eeq
Using the integration by parts (IBP) identities
\bea \label{IBP}
0&\sim & \frac{d}{dT}\left(\frac{e^{iuT}}{(1+T^2)^\alpha}f(T) \right)\,,\nonumber\\
0&\sim & \frac{d}{dT}\left(\frac{T e^{iuT}}{(1+T^2)^\alpha}f(T) \right)\,,
\eea
we get
\bea
\label{rec_relgen}
0&=& iu I_\alpha^f-2\alpha \widehat T I_{\alpha+1}^f+ I_\alpha^{f'}
\,,\nonumber\\
0&=& (1+i u\widehat T-2\alpha)I_\alpha^f+2 \alpha I_{\alpha+1}^f+\widehat T  I_\alpha^{f'}
\,,
\eea
In the case where the derivative $f'(T)$ satisfies the relation \eq{f'} we get the following more informative
result
\bea
\label{rec_rel}
0&=& iu I_\alpha^f-2\alpha \widehat T I_{\alpha+1}^f+Q_{\alpha+\beta}
\,,\nonumber\\
0&=& (1+i u\widehat T-2\alpha)I_\alpha^f+2 \alpha I_{\alpha+1}^f+\widehat T Q_{\alpha+\beta}
\,.
\eea
The latter result yields ladder relations expressing $I_\alpha^f$ in terms of 
first-order $u$-derivatives of any one of its two neighbors $ I_{\alpha+1}^f$ or  $ I_{\alpha-1}^f$,
and of extra source terms involving the  known, simpler integrals $Q_{\alpha+\beta}$, and their first-order $u$-derivative.

If we combine the general ladder relations \eq{rec_relgen} we get a second-order 
 inhomogeneous differential equation for $ I_\alpha^f$, namely
  \be \label{inhomBesselgen}
 (\widehat T^2+1 - \frac{2(\alpha-1)}{i u} \widehat T)I_\alpha^f(u)= -\frac1{iu} (\widehat T^2+1) I_\alpha^{f'}= -\frac1{iu}  I_{\alpha-1}^{f'} \,.
 \ee
In the particular case where  $f'(T)=\frac{1}{(1+T^2)^\beta}$, this second-order 
 inhomogeneous differential equation simplifies to
 \be \label{newBessel}
 (\widehat T^2+1 - \frac{2(\alpha-1)}{i u} \widehat T)I_\alpha^f(u)= -\frac1{iu} Q_{\alpha+\beta-1}\,.
 \ee

Especially useful is the second ladder relation in \eq{rec_rel}. Indeed, it can be solved for  $ I_{\alpha+1}^f$ 
when $\alpha \neq 0$, namely
\be \label{up}
 I_{\alpha+1}^f = \frac1{2\alpha} (2 \alpha-1 - iu \widehat T)  I_{\alpha}^f  -  \frac1{2\alpha} \widehat T Q_{\alpha+\beta}\,.
\ee
When $\alpha \neq0$, one can express, by using this relation recursively,  the infinite sequence of integrals $ I_{\alpha+n}^f $,
$n=0,1,2,3,\cdots$ in terms of the single integral $ I_{\alpha}^f $, and of derivatives of the simpler $Q_{\alpha+\beta}$'s . And when $\alpha = 0$, one can
express the  infinite sequence of integrals $ I_{n}^f $ in terms of only two master integrals  $ I_{0}^f $
and  $ I_{1}^f $ and the $Q_{\alpha+\beta}$'s. Moreover, when $\alpha=0$, one can use the first
ladder relation \eq{rec_rel} to relate  $ I_{0}^f $ to $Q_\beta$, namely
\be
 I_{0}^f (u) =  \frac{i}{u} Q_\beta(u)\,.
\ee
In other words, in all cases {\it only one new master integral}  is needed: $ I_{\alpha}^f $ if $\alpha\neq0$,
and   $ I_{1}^f $ if $\alpha=0$. 

Eqs. \eqref{rec_rel} and \eqref{newBessel} can be cast in the following compact form
\bea
\begin{pmatrix}
I_{\alpha+1}^f \cr
\frac{d}{du} I_{\alpha+1}^f 
\end{pmatrix}
={\mathcal A}(\alpha) \begin{pmatrix}
I_{\alpha}^f \cr
\frac{d}{du} I_{\alpha}^f 
\end{pmatrix}
+{\mathcal B}(\alpha) \,,
\eea
where ${\mathcal A}(\alpha)$ is a 2$\times$2 matrix with elements
\bea
{\mathcal A}_{11}(\alpha)&=& 1-\frac{1}{2\alpha}\,,\nonumber\\
{\mathcal A}_{12}(\alpha)&=&{\mathcal A}_{21}(\alpha) =-\frac{u}{2\alpha}\,,\nonumber\\
{\mathcal A}_{22(\alpha)}&=&0\,, 
\eea
and where the column matrix ${\mathcal B}(\alpha)$ reads
\bea
{\mathcal B}_{1}(\alpha)&=& \frac{i}{2\alpha}\frac{d}{du}Q_{\alpha+\beta}(u)\,,\nonumber\\
{\mathcal B}_{2}(\alpha)&=& \frac{i}{2\alpha}\left(\frac{d^2}{du^2}Q_{\alpha+\beta}(u) +Q_{\alpha+\beta-1}(u)\right)\,.
\eea
Recalling the definition \eqref{Q}, the last term ${\mathcal B}_{2}(\alpha)$ can also be rewritten as
\beq
{\mathcal B}_{2}(\alpha)= \frac{i}{2\alpha} Q_{\alpha+\beta}(u)\,,
\eeq
so that
\beq
{\mathcal B}_{1}(\alpha)=\frac{d}{du}{\mathcal B}_{2}(\alpha) \,.
\eeq

The first two sequences entering the two-loop waveform are:
(i)  $Q_{n+\frac12}^{\rm at}(u)$ (with $\beta=1$) which can be reduced to $Q_{\frac12}^{\rm at}(u)$
and combination of derivatives of $Q_{n+\frac12}(u)$, further reducible to $K_0$ and $K_1$;
and (ii)  $Q_{n}^{\rm as}(u)$ (with $\beta=\frac12$) which can be reduced to $Q_{1}^{\rm as}(u)$, its derivative, and to 
the $Q_{n+\frac12}(u)$'s, i.e., again to $K_0$ and $K_1$. 

The third sequence $Q_{n +\frac12}^{{\rm as}2}(u)$ is more complicated because $f'(T)$ is not of the form
$\frac{1}{(1+T^2)^\beta}$. However, it is of the form
\be \label{das2}
\frac{d}{dT} {\rm arcsinh}^2(T)= 2 \frac{{\rm arcsinh}(T)}{\sqrt{1+T^2}}\,.
\ee
In that case we get ladder relations with  source terms of the form of the already reduced $Q_{n}^{\rm as}(u)$'s,
modulo a single new master integral $Q_{\frac12}^{{\rm as}2}(u)$. We have shown here how to express the waveform in terms of  a small set of master integrals,
namely $Q_{\frac12}^{\rm at}(u)$, $Q_{1}^{\rm as}(u)$ and $Q_{\frac12}^{{\rm as}2}(u)$.
We can then consider that the latter functions are defined by their integral representations.
We can also give alternative representations of these master integrals.
In particular, we can use the inhomogeneous
second-order differential equation they satisfy. The left-hand side (lhs) of Eq. \eq{inhomBesselgen} features
an operator equivalent to the Bessel operator (if one replaces $I_\alpha^f(u) \to  u^{-\frac{1}{2}+\alpha} F(u)$,
and $\alpha \to  \nu + \frac1{2} $), say
\beq
{\mathcal L}_\nu (F(u))=\left(u^2\frac{d^2}{du^2}  +u \frac{d}{du} -(u^2+\nu^2)\right)F(u)=0\,,
\eeq 
 with solutions $ I_\nu(u)$,  and $K_\nu(u)$.
 This allows one to solve an inhomogeneous equation of the type
 \be
 {\mathcal L}_\nu (F(u))= S(u)\,,
 \ee
 with the boundary condition $F(u) \to 0$ as $u \to +\infty$, by using a Green's function bilinear in  $ I_\nu(u)$,  and $K_\nu(u)$ (or, equivalently, by using 
 Lagrange's method of varying constants). At the end, the solution can be expressed in terms of
 indefinite integrals $\int du  I_\nu(u) S(u)$ and  $\int du  K_\nu(u) S(u)$, with source terms $S(u)$ given
by modified Bessel functions.

We find that our first two master integrals satisfy the equations
\be
{\mathcal L}_0( Q_{\frac12}^{\rm at}(u))=-2iuK_0(u)\,.
\ee
and
\beq
\label{eq_Qas1}
\frac{d^2}{du^2} Q_{1}^{\rm as}(u) - Q_{1}^{\rm as}(u)=-\frac{2 i}{u}K_0(u)\,.
\eeq
The differential operator appearing in the latter equation is equivalent to the Bessel operator ${\mathcal L}_{\frac12}$ applied to $u^{-1/2} Q_{1}^{\rm as}(u)$, that is Eq. \eqref{eq_Qas1} becomes
\be
{\mathcal L}_{\frac12}( u^{-1/2}Q_{1}^{\rm as}(u))=-2iu^{1/2}K_0(u)\,.
\ee
This approach also applies to $Q_{\frac12}^{\rm as2}(u)$. Indeed, the rhs of Eq. \eq{inhomBesselgen}
features (when $\alpha=\frac12$, taking into account Eq. \eq{das2}) the function $Q^{\rm as}_{\alpha-\frac12}= Q^{\rm as}_{0}$. However, the first ladder relation \eq{rec_rel} allows one, when $\alpha=0$, to express $ Q^{\rm as}_{0}(u)$ in terms
of $K_0(u)$. We thereby obtain the following inhomogeneous Bessel equation for  $Q_{\frac12}^{\rm as2}(u)$ 
\be
{\mathcal L}_{0} Q_{\frac12}^{\rm as2}(u) = 4 K_0(u)\,.
\ee

One can easily recognize in the latter inhomogeneous Bessel equation the result of differentiating twice
the Bessel equation ${\mathcal L}_{\nu} K_\nu(u)$ with respect to the order $\nu$, before setting $\nu \to 0$.
One can then show (see Appendix \ref{eval_int} for details) that $Q_{\frac12}^{{\rm as}2}(u)$ can be expressed in terms
of the second $\nu$-derivative of $K_\nu(u)$ at $\nu=0$, through
\beq
\label{Qas2_1_2bis}
Q_{\frac12}^{\rm as2}(u)=-\frac{\pi^2}{2}K_0(u)+2\frac{d^2}{d\nu^2} K_\nu(u)\bigg|_{\nu=0}\,.
\eeq
We can then use the literature on second $\nu$-derivatives of $K_\nu(u)$ (see Refs. \cite{Brychkov:2016,NIST})  
to further  express $Q_{\frac12}^{\rm as2}(u)$ in terms of Meijer G functions. 
We also show in the Appendices how to express the two other master integrals in terms of Meijer G functions.

The main purpose of the present Section was to indicate the structures (together with simple IBP's)
 allowing one to reduce the computation of the 4PM waveform to a small number of master integrals.
 We leave to the future the issue of finding the most useful (from a practical point of view)
 master integrals. For instance, it may happen that the reduction of $Q_{\frac12}^{\rm as2}(u)$ 
 to Meijer G functions is less useful than considering simply that the function $Q_{\frac12}^{\rm as2}(u)$
 is defined by a specific integral (which can be numerically evaluated when needed).
 
 More details of the explicit reduction process we used are given in Appendix \ref{eval_int}.

\section{Explicit expression for the  Newtonian, 1PN and 2PN contributions to the 4PM waveform}

Considering, for simplicity,  only the $\ell m=22$ SWSH contribution to $U_2^{G^3}$, at PN orders $\eta^0$,
$\eta^2$, and $\eta^4$ we have
 four different types of contributions: arctan ($p_{22}^{Q^{\rm at}}$), arcsinh ($p_{22}^{Q^{\rm as}}$), arcsinh$^2$ ($p_{22}^{Q^{{\rm as}2}}$), together with the simpler type involving $Q_{\alpha}$ integrals ($p_{22}^Q$).
For example, we formally write, at each considered PN order, 
\bea
U_{22}^{G^3\eta^n}&=&\frac{\nu G^3 M^4 }{b^2 p_\infty^{5-n}}\left(p^{22 \eta^n}_{Q^{\rm at}}+p^{22 \eta^n}_{Q^{\rm as}}+p^{22 \eta^n}_{Q^{{\rm as}2}}+p^{22 \eta^n}_{Q} \right)\,.\nonumber\\
\eea
All the contributions $U_{2m}^{G^3\eta^n}$ are given in the companion ancillary file.
Explicitly, upon reduction to master integrals,  $U_{22}^{G^3\eta^0}$ reads
\begin{widetext}
\bea
U_{22}^{G^3\eta^0}&=&\frac{\nu G^3 M^4 \sqrt{5\pi}}{5   b^2 p_\infty^5} 
 \left[2 \left( u^2 + 3 u  + 2\right) K_0(u)+
  \left(2 u^2 + 5u  + 4 \right) K_1(u)-
u^2  \left(u +1\right)\frac{d}{du}Q_{\frac12}^{\rm as2}(u) +
 u^2\left( u + \frac{1}{2} \right) Q_{\frac12}^{\rm as2}(u)\right]
\,,\nonumber\\
U_{22}^{G^3\eta^2}&=&\frac{\nu G^3 M^4 \sqrt{5\pi}}{u^2  b^2 p_\infty^3} 
 \left[ \left( \left(\frac{12 \nu }{35}-\frac{19}{105}\right) u^5+\left(\frac{10 \nu }{7}-\frac{51}{14}\right)
   u^4+\left(\frac{18 \nu }{7}+\frac{337}{105}\right) u^3+\left(\frac{44 \nu }{35}+\frac{228}{35}\right) u^2\right) K_0(u)\right.\nonumber\\  
&+&  \left(\left(\frac{12
   \nu }{35}-\frac{19}{105}\right) u^5+\left(\frac{44 \nu }{35}-\frac{373}{105}\right) u^4+\left(\frac{79 \nu }{35}+\frac{61}{105}\right)
   u^3+\left(\frac{44 \nu }{35}-\frac{2}{7}\right) u^2  \right) K_1(u)\nonumber\\
&+&   \left(\left(\frac{19}{210}-\frac{6 \nu
   }{35}\right) u^6+\left(\frac{269}{420}-\frac{13 \nu }{35}\right) u^5+\left(\frac{19}{70}-\frac{11 \nu }{35}\right) u^4 \right)\frac{d}{du}Q_{\frac12}^{\rm as2}(u) \nonumber\\ 
&+&  \left( 
\left(\frac{6 \nu }{35}-\frac{19}{210}\right) u^6+\left(\frac{2 \nu }{7}-\frac{25}{42}\right)
  u^5+\left(\frac{11 \nu }{70}-\frac{17}{35}\right) u^4 \right) Q_{\frac12}^{\rm as2}(u)\nonumber\\
&+& \left(\frac{6 i u^4}{5}+\frac{6 i u^3}{5}+\frac{6 i u^2}{5}\right) Q_{1}^{\rm as}(u)+ \left(-\frac{6 i  u^4}{5}-\frac{6 i u^3}{5}\right) \frac{d}{du}Q_{1}^{\rm as}(u)\nonumber\\
&+&\left.  \left(\frac{12 i u^3}{5}+\frac{6 i u^2}{5}\right) Q_{\frac12}^{\rm at}(u)+ \left(-\frac{12 i u^3}{5}-\frac{12 i u^2}{5}\right) \frac{d}{du}Q_{\frac12}^{\rm at}(u)
\right]\,,
\eea

\bea
U_{22}^{G^3\eta^4}&=& \frac{\nu G^3 M^4 \sqrt{5\pi}}{u^2  b^2 p_\infty} 
 \left[ \left( \left(\frac{82
   \nu ^2}{945}-\frac{247 \nu }{945}+\frac{71}{945}\right) u^6+\left(\frac{262 \nu ^2}{945}-\frac{3433 \nu }{945}+\frac{4309}{1890}\right)
   u^5\right.\right.\nonumber\\
&+&\left.\left(\frac{136 \nu ^2}{189}+\frac{4867 \nu }{945}-\frac{12101}{945}\right) u^4+\left(\frac{256 \nu ^2}{315}+\frac{6361 \nu
   }{630}-\frac{1903}{180}\right) u^3+\left(\frac{32 \nu ^2}{105}+\frac{787 \nu }{105}+\frac{1138}{105}\right) u^2 
\right) K_0(u) \nonumber\\  
&+&  \left(\left(\frac{82 \nu ^2}{945}-\frac{247 \nu }{945}+\frac{71}{945}\right) u^6+\left(\frac{221 \nu ^2}{945}-\frac{6619 \nu
   }{1890}+\frac{2119}{945}\right) u^5+\left(\frac{596 \nu ^2}{945}+\frac{7157 \nu }{1890}-\frac{45053}{3780}\right) u^4
\right.\nonumber\\
&+&\left. \left(\frac{232 \nu
   ^2}{315}+\frac{9641 \nu }{1260}-\frac{2092}{315}\right) u^3+\left(\frac{32 \nu ^2}{105}+\frac{589 \nu }{105}-\frac{379}{30}\right)
   u^2 \right) K_1(u)\nonumber\\
&+&   \left(\left(-\frac{41 \nu ^2}{945}+\frac{247 \nu }{1890}-\frac{71}{1890}\right) u^7+\left(-\frac{7 \nu
   ^2}{135}+\frac{1157 \nu }{1890}-\frac{1073}{3780}\right) u^6+\left(-\frac{2 \nu ^2}{21}+\frac{37 \nu }{30}-\frac{19}{84}\right)
   u^5\right.\nonumber\\
&+&\left. \left(-\frac{8 \nu ^2}{105}+\frac{43 \nu }{84}+\frac{913}{840}\right) u^4  \right)\frac{d}{du}Q_{\frac12}^{\rm as2}(u) \nonumber\\ 
&+&  \left(\left(\frac{41 \nu ^2}{945}-\frac{247 \nu }{1890}+\frac{71}{1890}\right) u^7+\left(\frac{19 \nu
   ^2}{630}-\frac{689 \nu }{1260}+\frac{167}{630}\right) u^6+\left(\frac{16 \nu ^2}{315}-\frac{1117 \nu }{1260}-\frac{79}{504}\right)
   u^5\right.\nonumber\\
&+&\left. \left(\frac{4 \nu ^2}{105}-\frac{677 \nu }{840}+\frac{109}{210}\right) u^4 
 \right) Q_{\frac12}^{\rm as2}(u)\nonumber\\
&+& \left(
\left(\frac{39 i \nu }{35}-\frac{61 i}{70}\right) u^5+\left(\frac{54 i \nu }{35}-\frac{67
   i}{35}\right) u^4+\left(\frac{54 i \nu }{35}-\frac{67 i}{35}\right) u^3+\left(\frac{9 i \nu }{7}-\frac{51 i}{70}\right) u^2 \right) Q_{1}^{\rm as}(u)\nonumber\\
&+& \left( 
\left(\frac{61 i}{70}-\frac{39 i \nu }{35}\right) u^5+\left(\frac{67 i}{35}-\frac{54 i \nu }{35}\right)
   u^4+\left(\frac{51 i}{70}-\frac{9 i \nu }{7}\right) u^3\right) \frac{d}{du}Q_{1}^{\rm as}(u)\nonumber\\
&+&  \left(\left(\frac{114 i \nu }{35}-\frac{143 i}{35}\right)
   u^4+\left(\frac{99 i \nu }{35}+\frac{109 i}{70}\right) u^3+\left(\frac{9 i \nu }{7}-\frac{51 i}{70}\right) u^2 \right) Q_{\frac12}^{\rm at}(u)\nonumber\\
&+&\left. \left(\left(\frac{143 i}{35}-\frac{114 i \nu }{35}\right) u^4+\left(\frac{17 i}{35}-\frac{156 i \nu
   }{35}\right) u^3+\left(-\frac{18 i \nu }{7}-\frac{243 i}{35}\right) u^2 \right) \frac{d}{du}Q_{\frac12}^{\rm at}(u)
\right]\,, 
\eea

\end{widetext}
Reconstructing the original quantity $U_2^{G^3}$ is then straightforward
\bea
U_2^{G^3} &=& (U_{22}^{G^3\eta^0} + U_{22}^{G^3\eta^2}\eta^2 + U_{22}^{G^3\eta^4}\eta^4)  Y_{\bar 2; 22}(\theta, \phi) \nonumber\\
&+&(U_{20}^{G^3\eta^0} + U_{20}^{G^3\eta^2}\eta^2 + U_{20}^{G^3\eta^4}\eta^4) Y_{\bar 2; 20}(\theta, \phi) \nonumber\\
&+& (U_{2\bar 2}^{G^3\eta^0} + U_{2\bar 2}^{G^3\eta^2}\eta^2 + U_{2\bar 2}^{G^3\eta^4}\eta^4)  Y_{\bar 2; 2\bar 2}(\theta, \phi) \,,\nonumber\\
\eea
where the notation $\bar 2=-2$ has been used.

For instance, the Newtonian term $U_2^{G^3\eta^0}$ evaluated at $\theta=\frac{\pi}{2}$ is given by
\bea
U_2^{G^3\eta^0}&=&\frac{\nu G^3 M^4}{b^2 p_\infty^5} 
 \left[\right.\nonumber\\
&&\left(\left(\frac{u^2}{2}+1\right) \cos (2 \phi )+\frac{3}{2} i u \sin (2 \phi )\right) K_0(u) \nonumber\\
&+&
  \left(\left(\frac{i
   u^2}{2}+i\right) \sin (2 \phi )+\frac{5}{4} u \cos (2 \phi )+\frac{u}{4}\right) K_1(u)\nonumber\\
&+&
\left(-\frac{1}{4} u^3 \cos (2 \phi )-\frac{1}{4} i u^2 \sin (2 \phi
   )\right)\frac{d}{du}Q_{\frac12}^{\rm as2}(u) \nonumber\\
&+&\left.
\left(\frac{1}{4} i u^3 \sin (2 \phi )+\frac{1}{8} u^2 \cos (2 \phi
   )+\frac{u^2}{8}\right) Q_{\frac12}^{\rm as2}(u)\right]
\,,\nonumber\\
\eea
with (see Eq. \eqref{Qas2_1_2})
\bea
Q_{\frac12}^{\rm as2}(u)&=&-\frac{\pi^2}{2}K_0(u)+2\frac{d^2}{d\nu^2} K_\nu(u)\bigg|_{\nu=0}
\,,\nonumber\\
\frac{d}{du}Q_{\frac12}^{\rm as2}(u)&=&\frac{\pi^2}{2}K_1(u)-2\frac{d^2}{d\nu^2} K_\nu(u)\bigg|_{\nu=1}\,.
\eea

\section{Soft expansion of the 4PM Waveform}

The first three terms of the  low frequency (\lq\lq soft") expansion of the complex waveform  (in its rescaled version 
$W (\omega,{\bf n}) = \frac{c^4}{4 G}h_c(\omega,{\bf n})$) are universal
and have  the structure
\be\label{soft0}
 W(\om) \overset{\om \to 0^+}{=} \frac{{\mathcal A} }{\om} + {\mathcal B}  \ln  \om + {\mathcal C} \om (\ln \om)^2  
+ \cdots
 \ee
The leading-order term $\frac{{\mathcal A} }{\om}$ was derived in the classic work of Weinberg \cite{Weinberg:1965nx}.
The sub-leading  soft terms ${\mathcal B}  \ln  \om$ and  ${\mathcal C} \om (\ln \om)^2$ were derived in Refs. \cite{Saha:2019tub,Sahoo:2020ryf,Sahoo:2021ctw}, and have been re-written in a simplified way in Appendix B of Ref. \cite{Bini:2023fiz} in the case of the 
conservative $2 \to 2$ scattering, in the cm frame. [Radiative contributions to the scattering angle start at order $O\left(\frac{G^3}{c^5}\right)$ while the present study is limited to the $O\left(\frac{G^3}{c^4}\right)$ accuracy.]
We have found that the results obtained from these universal soft theorems agree with the direct low-frequency expansion of our explicit frequency-domain waveform presented above,  
 which read (limiting our computations to the equatorial plane, and including in the waveform only even-in-$\phi$ terms, for simplicity)
\begin{widetext}
 \bea
 {\mathcal A}^{G^3}  &=& \frac {i M^4 G^3 \nu} {b^3 p_\infty^4}  \left \{
\sin (2 \phi )+\eta ^2
        p_\infty^2 \left[\left (2 \nu - \frac {1} {2} \right) \sin(2 \phi ) + \left (\frac {9 \nu } {2} - \frac {3} {2} \right) \sin(4 \phi ) \right]\right.\nonumber\\
&+&\left.
\eta ^4 p_\infty^4 \left[\left (\frac {11 \nu ^2} {16} + \frac {105  \nu} {16} - \frac {251} {16} \right) \sin (2 \phi ) +  
\left (\frac {21 \nu ^2} {4} + \frac {45 \nu } {4} - \frac {9} {2} 
\right) \sin(4\phi ) + \left (\frac {95 \nu ^2} {16} - \frac {95 \nu} {16} + \frac {19} {16} \right) \sin(6 \phi ) \right]\right\}, \nn \\
  {\mathcal B}^{G^3}  &=& \frac{M^4 G^3 \nu}{b^2 p_\infty^5}  \left\{-\cos (2 \phi )+\eta ^2 p_\infty^2 \left[(-2 \nu -2) \cos (2 \phi )+(1-3 \nu ) \cos (4 \phi)\right]\right.\nonumber\\
&+&\left.
\eta ^4 p_\infty^4 \left[\left(-\frac{11 \nu ^2}{16}-\frac{57 \nu }{16}+\frac{65}{16}\right) \cos (2\phi )+\left(-\frac{7 \nu ^2}{2}-3 \nu +\frac{3}{2}\right) \cos (4 \phi )+\left(-\frac{45 \nu ^2}{16}+\frac{45 \nu}{16}-\frac{9}{16}\right) \cos (6 \phi )\right]
\right\} , \nn \\
   {\mathcal C}^{G^3}  &=&  \frac{i M^4 G^3 \nu}{2 b p_\infty^6}  \left\{ \sin (2 \phi )+\eta ^2 p_\infty^2 \left(\left(2 \nu -\frac{3}{2}\right) \sin (2 \phi
   )+\left(\frac{3 \nu }{2}-\frac{1}{2}\right) \sin (4 \phi )\right)\right.\nonumber\\
&+&\left.
\eta ^4 p_\infty^4 \left(\left(\frac{11 \nu ^2}{16}-\frac{55 \nu }{16}-\frac{83}{16}\right) \sin (2
   \phi )+\left(\frac{7 \nu^2}{4}-\frac{15 \nu }{4}+1\right) \sin (4 \phi )+\left(\frac{15 \nu ^2}{16}-\frac{15 \nu
   }{16}+\frac{3}{16}\right) \sin (6 \phi )\right)\right\}\,.
   \eea
\end{widetext}  
The latter 2PN-accurate expansions were obtained by deriving the low-frequency expansion ($u\to 0$) of our master integrals  $Q^{\rm at}_{\frac12}(u)$, $Q^{\rm as}_{1}(u)$ and $Q^{\rm as2}_{\frac12}(u)$ from their corresponding Meijer G representations, see Eqs. \eqref{eqA15} and   \eqref{Qas1smallu}-\eqref{Qas2smallu} below.

\section{Concluding remarks}

As a benchmark in the current effort to increase the PM accuracy  in solving the gravitational bremsstrahlung problem, we have provided here the  $O(G^4)$ (two-loop) asymptotic waveform, at 2PN accuracy $c^4 h_c\sim O(G^4\eta^4)$, computing the multipolar contributions $U_2$, $V_2$, $U_3$, $V_3$, $U_4$, $V_4$, $U_5$, $V_5$ and $U_6$ at the needed accuracy level.
The resulting expressions in Fourier space have been found to be  more involved than at the previously treated $O(G^3)$ 
(one-loop) case. They involve a finite number of new master integrals of the form
displayed in Eqs. \eq{all_integrals}. We have explicitly dicussed the IBP reduction of these integrals, and shown
how they can also be expressed either as integrals of integrands bilinear in Bessel functions (of integer or half-integer, orders), or, more explicitly, in terms of MeijerG functions.
See Appendices below for details. The  low-frequency limit of our 2PN-accurate $O(G^4)$ waveform was
checked against soft theorems.

Finally, as a side study of our previous results on the $O(G^2)$ bremsstrahlung waveform, we computed
the $O(G^3)$ spectral densities of radiated energy and momentum (in the rest frame of particle $A$)
at the thirtieth order in velocity. These results are given here as additional benchmarks in the study
of the gravitational bremsstrahlung problem in the post-Minkowskian approach.

Our results are explicitly given in electronic form in the  ancillary files  of the arxiv submission  of this work.

\section*{Acknowledgments}
The present research was partially supported by the 2021
Balzan Prize for Gravitation: Physical and Astrophysical Aspects, awarded to T. Damour.
D.B. thanks the Institut des Hautes \'Etudes Scientifiques
for warm hospitality at various stages during the development of the present project and also 
acknowledges sponsorship of the Italian Gruppo Nazionale per la
Fisica Matematica (GNFM) of the Istituto Nazionale di
Alta Matematica (INDAM).  

\appendix 

\section{Evaluation, and reduction, of the integrals entering the $O(G^4)$ waveform}
\label{eval_int}

Let us discuss how to evaluate (and reduce) the needed integrals \eqref{all_integrals}, which are of the general form
\beq
I_\alpha^f(u)=\int dT \frac{e^{iu T}}{(1+T^2)^{\alpha}}f(T)\,,
\eeq
with $f(T)=[{\rm arctan}\left(T\right),{\rm arcsinh}(T),{\rm arcsinh}^2(T)]$. The first two are purely imaginary (for symmetry reasons) while the third one is real.
Let us discuss the first two types, having the special property that in both cases 
\beq
f'(T)=\frac{1}{(1+T^2)^\beta}\,,
\eeq
with $\beta=1$ for the arctan integral and $\beta=1/2$ for the arcsinh integral.
Denoting
\beq
\widehat T =\frac{1}{i}\frac{d}{du}\,,
\eeq
we have
\beq
\label{prima_rel}
(\widehat T^2+1)I_\alpha^f(u)=I_{\alpha-1}^f(u)\,.
\eeq
Moreover, the IBP relations, Eqs. \eq{IBP}, imply ($f(T)$ carrying the index $\beta$)
\bea
\label{rec_relapp}
0&=& iu I_\alpha^f-2\alpha \widehat T I_{\alpha+1}^f+Q_{\alpha+\beta}
\,,\nonumber\\
0&=& (1+i u\widehat T-2\alpha)I_\alpha^f+2 \alpha I_{\alpha+1}^f+\widehat T Q_{\alpha+\beta}
\,,
\eea
leading to the following recurrence relations
\bea
\label{rec_rel2}
I_{\alpha-1}^f&=& \frac{2(\alpha-1)}{iu}\widehat T I_\alpha^f -\frac{1}{iu}Q_{\alpha+\beta-1}
\,,\nonumber\\
I_{\alpha+1}^f&=& \frac{1}{2\alpha}(2\alpha-1-iu \widehat T )I_\alpha^f-\frac{1}{2\alpha}\widehat T Q_{\alpha+\beta}.
\eea
Therefore, as explained in the text, one can reduce the first two types of integrals to those with the lowest possible values of $\alpha$, namely $\alpha=\frac12$ for  $f(T)={\rm arctan}\left(T\right)$, and
$\alpha=1$ for  $f(T)={\rm arcsinh}(T)$. Some details follow.
.

\subsection{Integrals of the type $Q_{\alpha}^{\rm at}(u)$}

Let us consider the case $f(T)={\rm arctan}(T)$, i.e., $I_\alpha^f=Q_\alpha^{\rm at}$, $\beta=1$. 
The first equation of Eqs. \eqref{rec_rel} for the first few values of $\alpha=\frac12,\frac32,\frac52$ gives
\bea
R_{\frac12}\qquad0&=&iu Q_{\frac12}^{\rm at}(u)  + i\frac{d}{du}Q_{\frac32}^{\rm at}(u) + 2 uK_1(u) 
\,,\nonumber\\
R_{\frac32}\qquad0&=&iu Q_{\frac32}^{\rm at}(u) + 3i\frac{d}{du}Q_{\frac52}^{\rm at}(u)  + \frac{2}{3}u^2K_0(u)\nonumber\\
&+&  \frac{4}{3} uK_1(u) 
\,,\nonumber\\
R_{\frac52}\qquad0&=&iu Q_{\frac52}^{\rm at}(u)+5 i \frac{d}{du} Q_{\frac72}^{\rm at}(u) +  \frac{8}{15} u^2 K_0(u)\nonumber\\
&+& \frac{2u}{15}( u^2 + 8)K_1(u) 
\,,
\eea
while the second one gives
\bea
S_{\frac12}\qquad0&=&u \frac{d}{du}Q_{\frac12}^{\rm at}(u) + 2iu K_0(u)  +Q_{\frac32}^{\rm at}(u)
\,,\nonumber\\
S_{\frac32}\qquad0&=& -2Q_{\frac32}^{\rm at}(u) +u \frac{d}{du}Q_{\frac32}^{\rm at}(u)   + \frac23 iu^2 K_1(u) \nonumber\\
&+& 3Q_{\frac52}^{\rm at}(u)
\,,\nonumber\\
S_{\frac52}\qquad0&=& -4Q_{\frac52}^{\rm at}(u) +u \frac{d}{du}Q_{\frac52}^{\rm at}(u)  + \frac{2iu^3}{15} K_0(u) \nonumber\\
&+& \frac{4iu^2}{15}K_1(u) + 5Q_{\frac72}^{\rm at}(u)
\,.
\eea
Solving the relation $S_{\frac12}$ for $Q_{\frac32}^{\rm at}(u)$ and substituting it into the relation $R_{\frac12}$ gives the following  second order differential equation for $Q_{\frac12}^{\rm at}(u)$
\beq
\label{rel_Q32}
 -iu\frac{d^2}{du^2}Q_{\frac12}^{\rm at}(u) -i\frac{d}{du}Q_{\frac12}^{\rm at}(u)+ iuQ_{\frac12}^{\rm at}(u)+2K_0(u)=0\,.
\eeq
The latter is a real equation since $Q_{\frac12}^{\rm at}(u)=i \widetilde Q_{\frac12}^{\rm at}(u)$, with $\widetilde Q_{\frac12}^{\rm at}(u)$ a real function of $u$, so that 
\beq
u\frac{d^2}{du^2}\widetilde Q_{\frac12}^{\rm at}(u) +\frac{d}{du}\widetilde Q_{\frac12}^{\rm at}(u)-u\widetilde Q_{\frac12}^{\rm at}(u)=-2K_0(u)\,.
\eeq
Recalling Bessel's modified differential equation of order $n$
\beq
{\mathcal L}_n (f(u))=\left(u^2\frac{d^2}{du^2}  +u \frac{d}{du} -(u^2+n^2)\right)f(u)=0\,,
\eeq 
with solutions $f_n(u)=I_n(u),K_n(u)$, 
the above equation reads
\beq
{\mathcal L}_0(\widetilde Q_{\frac12}^{\rm at}(u))=-2uK_0(u)\,.
\eeq
The solution for $\widetilde Q_{\frac12}^{\rm at}(u)$ is thus given by
\bea
\label{sol_tildeQ1}
\widetilde Q_{\frac12}^{\rm at}(u)&=&c_1K_0(u)+c_2I_0(u)
-2I_0(u)\int_0^u K^2_0(x)dx \nonumber\\ 
&+& 2K_0(u)\int_0^u I_0(x)K_0(x) dx  \,,
\eea
namely $\widetilde Q_{\frac12}^{\rm at}(u)$ is expressed as some kind of {\it iterated Bessel function}.
The integration constants $c_1$ and $c_2$ corresponding to the solutions of the associated homogeneous equation are chosen in order to ensure that
$\widetilde Q_{\frac12}^{\rm at}(u)$ satisfies the needed limits as $u\to0$  and $u\to\infty$.
This implies $c_1=0$ and $c_2=\frac{\pi^2}{2}$.
The final solution can be expressed in terms of Meijer G functions as follows
\bea
\label{meijer}
\widetilde Q_{\frac12}^{\rm at}(u)&=&\frac{\pi^2}{2}I_0(u) \nonumber\\
&-& 2I_0(u)\frac{\sqrt{\pi }u}{4}  G_{2,4}^{3,1}\left(u^2\bigg|
\begin{array}{c}
 \frac{1}{2},\frac{1}{2} \\
 0,0,0,-\frac{1}{2} \\
\end{array}
\right)\nonumber\\  
&+& 2K_0(u)\frac{u}{4 \sqrt{\pi }}G_{2,4}^{2,2}\left(u^2\bigg|
\begin{array}{c}
 \frac{1}{2},\frac{1}{2} \\
 0,0,-\frac{1}{2},0 \\
\end{array}
\right)\,.\nonumber\\
\eea

For small values of $u$ we find
\bea
\label{eqA15}
\widetilde Q_{\frac12}^{\rm at}(u)&=&
u\left(2+\frac{5 u^2}{18}+\frac{89 u^4}{7200}\right)\ln\left(\frac12 u e^\gamma\right)\nonumber\\
&+&
\frac{\pi ^2}{2}\left(1+\frac{ u^2}{4}+\frac{u^4}{64}\right)\nonumber\\
&-&u\left( 4+\frac{37}{54} u^2+\frac{7393}{216000} u^4\right)\nonumber\\
&+& O(u^6)\,,
\eea
whereas for large values of $u$
\bea
\widetilde Q_{\frac12}^{\rm at}(u)&\sim&\frac{\sqrt{2\pi}}{4u^{3/2}}e^{-u}\left[1+\left(2u - \frac14\right)\ln\left(8 u e^\gamma\right)\right]
\,.
\eea
Notice that this solution gives $\widetilde Q_{\frac12}^{\rm at}(0)=\frac{\pi^2}{2}$, as it can be checked directly by using the method of regions. Considering  complex values of $\alpha$, one can distinguish two regions:
 region 1 ($uT\ll1$), and region 2 ($uT\gg1$). One then formally power-expand the integrand in each region.
 In region 1 one gets zero, whereas in region 2, one gets $\pi^2/2$ upon changing to the variable $x=u T$ and then taking $\alpha \to 1/2$.
Beware that one cannot first take the limit $\alpha\to1/2$, and then take the limit $u\to0$ in the integrand because of 
the arising of a logarithmic divergence.

Using the recurrence relations above, we then find 
\begin{widetext}
\bea
Q_{\frac32}^{\rm at}(u) &=& -u \frac{d}{du}Q_{\frac12}^{\rm at}(u)  - 2i K_0(u) u
\,,\nonumber\\
Q_{\frac52}^{\rm at}(u) &=& - \frac{2}{3} u\frac{d}{du}Q_{\frac12}^{\rm at}(u)  - \frac{4}{3}i K_0(u) u  +\frac13 Q_{\frac12}^{\rm at}(u) u^2  - \frac{8}{9}i K_1(u) u^2 
\,,\nonumber\\
Q_{\frac72}^{\rm at}(u)  &=& \left(-\frac{16}{15}u  - \frac{46}{225}u^3 \right)iK_0(u) - \frac{64}{75} i K_1(u) u^2 \nonumber\\ 
&+& \left(-\frac{8}{15} u - \frac{1}{15} u^3\right)\frac{d}{du}Q_{\frac12}^{\rm at}(u) +  \frac{4}{15}Q_{\frac12}^{\rm at}(u) u^2\,,\nonumber\\
Q_{\frac92}^{\rm at}(u) &=& \left(-\frac{176}{735} u^3  - \frac{32}{35} u\right)iK_0(u) + \left(-\frac{192}{245} u^2  - \frac{352}{11025} u^4 \right)iK_1(u) \nonumber\\
&+& \left(-\frac{8}{105} u^3 - \frac{16}{35} u\right)\frac{d}{du}Q_{\frac12}^{\rm at}(u) + \left(\frac{8}{35} u^2 + \frac{1}{105} u^4\right)Q_{\frac12}^{\rm at}(u)
\,,\nonumber\\
Q_{\frac{11}2}^{\rm at}(u)&=& \left(-\frac{688}{2835} u^3  - \frac{256}{315} u  - \frac{1126}{297675} u^5 \right)iK_0(u) + \left(-\frac{2048}{2835} u^2  - \frac{2752}{59535} u^4 \right)iK_1(u)\nonumber\\ 
&+& \left(-\frac{8}{105} u^3 - \frac{128}{315} u - \frac{1}{945} u^5\right)\frac{d}{du}Q_{\frac12}^{\rm at}(u) + \left(\frac{4}{315} u^4 + \frac{64}{315} u^2\right)Q_{\frac12}^{\rm at}(u)\,.
\eea

\subsection{Integrals of the type $Q_{\alpha}^{\rm as}(u)$}

Let us consider now the case $f(T)={\rm arcsinh}(T)$, i.e., $I_\alpha^f=Q_\alpha^{\rm as}$, $\beta=1/2$. 
Using the recurrence relations \eqref{rec_rel} we find
\bea
Q_{2}^{\rm as}(u) &=&   \frac12 Q_{1}^{\rm as}(u)  - \frac{u}{2}  \frac{d}{du}Q_{1}^{\rm as}(u)  -i u K_0(u)\nonumber\\
Q_{3}^{\rm as}(u) &=&   - \frac{3}{8} u\frac{d}{du}Q_{1}^{\rm as}(u)   + \left(\frac{3}{8} + \frac{u^2}{8}\right)Q_{1}^{\rm as}(u) - \frac{3iu}{4} K_0(u) - \frac{5iu^2}{12}K_1(u) \nonumber\\
Q_{4}^{\rm as}(u) &=&+ \left(\frac{5}{16} + \frac{u^2}{8}\right)Q_{1}^{\rm as}(u)+ \left(-\frac{5}{16} u - \frac{1}{48} u^3\right)\frac{d}{du}Q_{1}^{\rm as}(u) \nonumber\\
&+& \left(-\frac{5 i u}{8} - \frac{11}{120} i u^3\right)K_0(u) - \frac{161}{360} i K_1(u) u^2   \nonumber\\
Q_{5}^{\rm as}(u) &=& \left(\frac{15}{128} u^2 + \frac{1}{384} u^4 + \frac{35}{128}\right)Q_{1}^{\rm as}(u) + \left(-\frac{5}{192} u^3 - \frac{35}{128} u\right)\frac{d}{du}Q_{1}^{\rm as}(u)\nonumber\\
&+&
\left(-\frac{35}{64} i u  - \frac{587}{5040} i u^3 \right)K_0(u) + \left(-\frac{969}{2240}i u^2  - \frac{31}{2240} i u^4 \right)K_1(u) 
 \nonumber\\
Q_{6}^{\rm as}(u) &=&+ \left(\frac{1}{256} u^4 + \frac{7}{64} u^2 + \frac{63}{256}\right)Q_{1}^{\rm as}(u)+ \left(-\frac{7}{256} u^3 - \frac{63}{256} u - \frac{1}{3840} u^5\right)\frac{d}{du}Q_{1}^{\rm as}(u) \nonumber\\ 
&+& \left(-\frac{63}{128} i u - \frac{24883}{201600} i u^3  - \frac{193}{120960}i u^5 \right)K_0(u) + \left(-\frac{82841}{201600} i u^2  - \frac{187}{8640} i u^4 \right)K_1(u) \,,
\eea
where $Q_{1}^{\rm as}(u)=i \widetilde Q_{1}^{\rm as}(u)$ satisfies the following (real) differential equation
\beq
\frac{d^2}{du^2}\widetilde Q_{1}^{\rm as}(u) -\widetilde Q_{1}^{\rm as}(u)=-\frac{2}{u}K_0(u)\,.
\eeq
The solution reads
\bea
\label{sol_tildeQ1as}
\widetilde Q_{1}^{\rm as}(u)&=&\frac{\pi^2}{2}e^{-u}
+\sqrt{\pi}e^{u}G_{2,3}^{3,0}\left(2 u\left|
\begin{array}{c}
 \frac{1}{2},1 \\
 0,0,0 \\
\end{array}
\right.\right)  
-\frac{e^{-u}}{\sqrt{\pi}}G_{2,3}^{3,1}\left(2 u\left|
\begin{array}{c}
 \frac{1}{2},1 \\
 0,0,0 \\
\end{array}
\right.\right)\,,
\eea
which vanishes both at $u=0$ and $u\to\infty$.

\subsection{Integrals of the type $Q_{\alpha}^{\rm as2}(u)$}
\label{int_arcsinhsquare}

Finally, let us consider the case $f(T)={\rm arcsinh}^2(T)$, i.e., $I_\alpha^f=Q_\alpha^{\rm as2}$, $\beta=1/2$. 
The basic integral with $\alpha=\frac12$ is 
\beq
\label{Qas212def}
Q_{\frac12}^{\rm as2}(u)=\int_{-\infty}^\infty dT \, \frac{e^{iuT}}{\sqrt{1+T^2}}{\rm arcsinh}^2 T\,,
\eeq
which can be computed as follows.
Consider the identity (\cite{NIST}, 10.32.7)
\beq
\cos (\frac{\nu}{2}\pi) K_\nu (x)=\int_0^\infty \cos(x\sinh t)\cosh (\nu t)dt=\frac12 \int_{-\infty}^\infty e^{i x\sinh t}\cosh (\nu t)dt\,.
\eeq
Differentiating both sides wrt $\nu$ one gets
\beq
-\frac{\pi}{2}\sin (\frac{\nu}{2}\pi) K_\nu (x)+\cos (\frac{\nu}{2}\pi)\frac{\partial}{\partial \nu} K_\nu (x)=
\frac12 \int_{-\infty}^\infty e^{i x\sinh t}t \sinh (\nu t)dt\,.
\eeq
Changing the variable as
\beq
T=\sinh(t)\,, \qquad dT=\cosh t dt\,,\qquad dt=\frac{dT}{\sqrt{1+T^2}}
\eeq
one gets
\beq
\label{Bessel_fund}
-\frac{\pi}{2}\sin (\frac{\nu}{2}\pi) K_\nu (x)+\cos (\frac{\nu}{2}\pi)\frac{\partial}{\partial \nu} K_\nu (x)=
\frac12 \int_{-\infty}^\infty e^{i x T}{\rm arcsinh}(T) \sinh (\nu {\rm arcsinh}(T)) \frac{dT}{\sqrt{1+T^2}}\,.
\eeq

This equation can be used to reduce the integrals
\be
I_3(n)=Q^{\rm as}_{n+\frac12}(u) = \int dT \frac{e^{iuT}}{(1+T^2)^{n+\frac12}}{\rm arcsinh} T
\ee
to  $K_n(u)$ Bessel functions   (and thereby to $K_0(u)$ and $K_1(u)$) and exponential functions. 
Indeed, taking $\nu=1$ in \eq{Bessel_fund} yields
\be
 \int_{-\infty}^\infty dT  e^{i x T}\frac{T}{\sqrt{1+T^2}} {\rm arcsinh}(T)= - \pi K_1(x)= + \pi K^{\prime}_0(x)\,.
\ee
Integrating both sides over $x$ then yields  
\be
Q^{\rm as}_{\frac12}(x)=  i \pi K_0(x)\,,
\ee
where both sides vanish as $x \to +\infty$.

Further differentiating both sides of \eq{Bessel_fund}  with respect to $\nu$ and evaluating for $\nu=0$ one finds
\beq
\label{caso_nu_eq_0}
-\frac{\pi^2}{4}K_0(x)+\frac{d^2}{d\nu^2} K_\nu(x)\bigg|_{\nu=0}=\frac12 \int_{-\infty}^\infty dT \, \frac{e^{ixT}}{\sqrt{1+T^2}}{\rm arcsinh}^2 T\,,
\eeq
implying 
\beq
\label{Qas2_1_2}
Q_{\frac12}^{\rm as2}(u)=-\frac{\pi^2}{2}K_0(u)+2\frac{d^2}{d\nu^2} K_\nu(u)\bigg|_{\nu=0}\,,
\eeq
where the second term is given in Eq. \eqref{dv2_besselK} below.

The simplest way to get the integrals corresponding to higher values of $\alpha$ consists in changing the variable as
\beq
T=\frac{v_0t}{b}\,,\qquad dT=\frac{v_0}{b}dt\,,
\eeq
differentiating then with respect to $b$ leading to
\bea
Q_{\frac32}^{\rm as2}(u)&=& -u\frac{dQ_{\frac12}^{\rm as2}(u)}{du}
- 2 \frac{d\widetilde Q_{1}^{\rm as}(u)}{du}\,,
\eea
where $\widetilde Q_{1}^{\rm as}(u)$ is given by Eq. \eqref{sol_tildeQ1as}.
Similarly we get
\bea
Q_{\frac52}^{\rm as2}(u)&=& -\frac23uK_1(u) 
+\frac13u^2 Q_{\frac12}^{\rm as2}(u)-\frac23u\frac{dQ_{\frac12}^{\rm as2}(u)}{du}
-\frac{4}{3} \frac{d\widetilde Q_{1}^{\rm as}(u)}{du} +u \widetilde Q_{1}^{\rm as}(u)\,,\nonumber\\
Q_{\frac72}^{\rm as2}(u)&=& -\frac{3}{10}u^2 K_0(u) -\frac23 u K_1(u)
+\left(-\frac{8}{15}u-\frac{1}{15}u^3 \right)\frac{dQ_{\frac12}^{\rm as2}(u)}{du}
+\frac{4}{15}u^2Q_{\frac12}^{\rm as2}(u)
\nonumber\\
&-&\left(\frac{u^2}{4}+\frac{16}{15}\right) \frac{d\widetilde Q_{1}^{\rm as}(u)}{du} +\frac{11}{12}u \widetilde Q_{1}^{\rm as}(u)\,,\nonumber\\
Q_{\frac92}^{\rm as2}(u)&=&  -\frac{13}{36}u^2 K_0(u) 
+ \left(-\frac{29}{420}u^3-\frac{28}{45}u \right)K_1(u)\nonumber\\ 
&+&\left(-\frac{16}{35}u-\frac{8}{105}u^3  \right)\frac{dQ_{\frac12}^{\rm as2}(u)}{du}
+\left(\frac{8}{35}u^2+\frac{1}{105}u^4 \right)Q_{\frac12}^{\rm as2}(u)
\nonumber\\
&-&\left(\frac{7}{24}u^2 +\frac{32}{35} \right) \frac{d\widetilde Q_{1}^{\rm as}(u)}{du} + \left(\frac{1}{24}u^3+\frac{33}{40}u \right)\widetilde Q_{1}^{\rm as}(u)\,,\nonumber\\
Q_{\frac{11}2}^{\rm as2}(u)&=&  \left(-\frac{65 u^4}{6048}-\frac{33647 u^2}{90720}  \right)K_0(u) 
+ \left(-\frac{3109 u^3}{30240}-\frac{328 u}{567}  \right)K_1(u)\nonumber\\ 
&+&\left( -\frac{u^5}{945}-\frac{8 u^3}{105}-\frac{128 u}{315} \right)\frac{dQ_{\frac12}^{\rm as2}(u)}{du}
+\left(\frac{4}{315}u^4 + \frac{64}{315}u^2  \right)Q_{\frac12}^{\rm as2}(u)
\nonumber\\
&-&\left( \frac{u^4}{192}+\frac{283 u^2}{960}+\frac{256}{315} \right) \frac{d\widetilde Q_{1}^{\rm as}(u)}{du} 
+ \left( \frac{17 u^3}{288}+\frac{5053 u}{6720} \right)\widetilde Q_{1}^{\rm as}(u)\,,\nonumber\\
Q_{\frac{13}2}^{\rm as2}(u)&=&  \left(
-\frac{109   u^4}{5940}-\frac{521129 u^2}{1425600}\right)K_0(u) 
+ \left(-\frac{281 u^5}{221760}-\frac{1188767 u^3}{9979200}-\frac{7664 u}{14175}\right)K_1(u)\nonumber\\ 
&+&\left(
-\frac{2 u^5}{1155}-\frac{256 u^3}{3465}-\frac{256 u}{693}
\right)\frac{dQ_{\frac12}^{\rm as2}(u)}{du}
+\left( 
\frac{u^6}{10395}+\frac{16 u^4}{1155}+\frac{128 u^2}{693}
\right)Q_{\frac12}^{\rm as2}(u)
\nonumber\\
&-&\left( 
\frac{5 u^4}{576}+\frac{1289 u^2}{4480}+\frac{512}{693}
\right) \frac{d\widetilde Q_{1}^{\rm as}(u)}{du} 
+ \left( 
\frac{u^5}{1920}+\frac{383 u^3}{5760}+\frac{5597 u}{8064}
\right)\widetilde Q_{1}^{\rm as}(u)\,. 
\eea

The remaining relations needed at 2PN can be found in the ancillary file.

\end{widetext}

\section{Derivatives of BesselK  functions with respect to the order: explicit expressions}

The first-order derivatives of the BesselK functions with respect to the order are given by
\bea
\frac{\partial}{\partial \nu}K_\nu(u)\bigg|_{\nu=0}&=&0\,,\nonumber\\
\frac{\partial}{\partial \nu}K_\nu(u)\bigg|_{\nu=1}&=&\frac{1}{u}K_0(u)\,.
\eea
The second-order derivatives are instead much more complicated (see e.g. Ref. \cite{Brychkov:2016}) and involve MeijerG functions. We find
\bea
\label{dv2_besselK}
\frac{\partial^2}{\partial \nu^2}K_\nu(u)\bigg|_{\nu=0}&=& A_{K_0}(u)K_0(u)+A_{I_0}(u)I_0(u)\,,\nonumber\\
\frac{\partial^2}{\partial \nu^2}K_\nu(u)\bigg|_{\nu=1}&=& B_{K_0}(u)K_0(u)+B_{K_1}(u)K_1(u)\nonumber\\
&+&B_{I_0}(u)I_0(u)+
B_{I_1}(u)I_1(u)\,,
\eea
where (using the traditional notation for the MeijerG functions)
\bea
A_{K_0}(u)&=&  -i\pi\ln\left(\frac{i}2 u e^{\gamma}\right)\nonumber\\
&-& \frac{i\pi}{4}u^2 {}_3F_4\left( 1, 1, \frac32; 2, 2, 2, 2; u^2\right)\nonumber\\
&-& \frac{\sqrt{\pi}}2  
G_{3,1}^{3,5}\left(-u^2 \bigg|
\begin{array}{c}
\frac12,-\frac12,  1 \\
0,0,0,0,-\frac{1}{2} \\
\end{array}
\right) \,,
\nonumber\\
A_{I_0}(u)&=&  -\frac{\pi^{3/2}}{2}
G_{4,5}^{4,0}\left(-u^2 \bigg|
\begin{array}{c}
0,\frac12, \frac12, 1 \\
0,0,0,0, \frac{1}{2} \\
\end{array}
\right) 
\nonumber\\
&+& \frac{i\pi^{3/2}}{2} 
G_{3,1}^{3,5}\left(-u^2 \bigg|
\begin{array}{c}
\frac12,-\frac12,  1 \\
0,0,0,0,-\frac{1}{2} \\
\end{array}
\right) \,,
\eea
and
\bea
B_{K_0}(u)&=&- \frac{i\pi}{u} -\frac{i\pi u }{2}\, {}_3F_4\left( 1, 1, \frac32; 2, 2, 2, 2; u^2\right)\nonumber\\
&-&\frac{3  i\pi u^3}{64}\, {}_3F_4\left( 2, 2, \frac52; 3, 3, 3, 3;u^2\right)\nonumber\\
&-&\frac{\sqrt{\pi}}{u} G_{2,4}^{2,1}\left(-u^2 \bigg|
\begin{array}{c}
\frac12, -\frac12 \\
0,0,0,-\frac{1}{2} \\
\end{array}
\right) \,,
\nonumber\\
B_{K_1}(u)&=&  \frac{i\pi}{2}\left[3i\pi +2\ln\left(\frac{u}{2}e^\gamma\right)\right] \nonumber\\
&+& \frac{i\pi u^2}{4}{}_3F_4\left( 1, 1, \frac32; 2, 2, 2, 2;u^2\right)\nonumber\\
&-& \frac{\sqrt{\pi}}{2} G_{2,5}^{3,1}\left(-u^2 \bigg|
\begin{array}{c}
\frac12, -\frac12, 1 \\
0,0,0,0,-\frac{1}{2} \\
\end{array}
\right) \,,
\nonumber\\
B_{I_0}(u)&=&  -\frac{\pi^{3/2}}{u}\left[ 
G_{3,4}^{3,0}\left(-u^2 \bigg|
\begin{array}{c}
0,\frac12, \frac12 \\
0,0,0,\frac{1}{2} \\
\end{array}
\right) 
\right. 
\nonumber\\
&+&\left. i 
G_{2,4}^{2,1}\left(-u^2 \bigg|
\begin{array}{c}
\frac12, -\frac12  \\
0,0,0,-\frac{1}{2} \\
\end{array}
\right) 
\right]\,,\nonumber\\
B_{I_1}(u)&=&  \frac{\pi^{3/2}}{u} \left[ 
G_{4,5}^{4,0}\left(-u^2 \bigg|
\begin{array}{c}
0,\frac12,  \frac12, 1  \\
0,0,0,0,\frac{1}{2} \\
\end{array}
\right) 
\right.\nonumber\\
&+&\left. i 
G_{3,5}^{3,1}\left(-u^2 \bigg|
\begin{array}{c}
\frac12,  -\frac12, 1  \\
0,0,0,0,-\frac{1}{2} \\
\end{array}
\right) 
\right]\,.
\eea

The following relation holds
\beq
\frac{d}{du}\left(\frac{\partial^2}{\partial \nu^2}K_\nu(u)\bigg|_{\nu=0}\right)=-\frac{\partial^2}{\partial \nu^2}K_\nu(u)\bigg|_{\nu=1}\,.
\eeq
  
Finally, using the results of this section, we can write the limiting expressions of the master integral for $u\to 0$:
\bea
\label{Qas1smallu}
Q_{1}^{\rm as}(u)&=& iu \left[c_1^{\rm as}+c_{\ln}^{\rm as}\ln(u)+ c_{\ln^2}^{\rm as}\ln^2u \right]+O(u^2)\,,\nonumber\\
\eea
with
\bea
c_1^{\rm as}&=& 2-2\gamma_E +\gamma_E^2-\frac{\pi^2}{6}+2\ln 2 -2\gamma_E \ln(2)+\ln^22\,,\nonumber\\
c_{\ln}^{\rm as}&=&-2+2\gamma_E-2\gamma_E \ln(2)\,, \nonumber\\
c_{\ln^2}^{\rm as}&=&1 \,.
\eea
Similarly,
\bea
\label{Qatsmallu}
Q_{\frac12}^{\rm at}(u)&=& i\frac{\pi^2}{2}+2iu \left[-2+ \gamma_E - \ln 2  + \ln (u)\right]\nonumber\\
&+& O(u^2)\,, 
\eea
and
\bea
\label{Qas2smallu}
Q_{\frac12}^{\rm as2}(u)&=&-\frac{2}{3} \log ^3(u)+\frac{1}{6} (12 \log (2)-12 \gamma_E ) \log ^2(u)\nonumber\\
&+&\frac{1}{6} \left(\pi ^2-12 \log ^2(2)+12
   \gamma_E  (\log (4)-\gamma_E )\right) \log (u)\nonumber\\
&+&\frac{1}{6} \left(
-8 \zeta (3)\right.\nonumber\\
&-&\left. \left((\gamma_E -\log (2)) \left((\log
   (4)-2 \gamma_E )^2-\pi ^2\right)\right)
\right)\nonumber\\
&+& O(u^2)\,. 
\eea

\section{Some properties of the Meijer-$G$ function: a brief reminder}

The Meijer-$G$ function is usually defined by the following Mellin-Barnes integral representation, see e.g., Eq. 16.17.1 of \cite{NIST}%
\begin{eqnarray}
&&G_{p,q}^{m,n}\left( z\left\vert 
\begin{array}{c}
a_{1},\ldots ,a_{p} \\ 
b_{1},\ldots ,b_{q}%
\end{array}%
\right. \right)  \label{Meijer_G_def} \\
&=&\frac{1}{2\pi i}\int_{L}\frac{\prod_{\ell =1}^{m}\Gamma \left( b_{\ell
}-s\right) \prod_{\ell =1}^{n}\Gamma \left( 1-a_{\ell }+s\right) }{%
\prod_{\ell =m}^{q-1}\Gamma \left( 1-b_{\ell +1}+s\right) \prod_{\ell
=n}^{p-1}\Gamma \left( a_{\ell +1}-s\right) }z^{s}ds,  \notag
\end{eqnarray}%
where the integration path $L$ separates the poles of the factors $\Gamma \left(
b_{\ell }-s\right) $ from those of the factors $\Gamma \left( 1-a_{\ell
}+s\right) $. Here, $m$ and $n$ are integers such that $0\leq m\leq q$ and $%
0\leq n\leq p$, and none of $a_{k}-b_{j}$ is a positive integer when $1\leq
k\leq n$ and $1\leq j\leq m$.

Let us shortly recall some properties of the Meijer-$G$ function (see, e.g., Section 8.2 of Ref. \cite{Prudnikov3}).
\begin{itemize}
  
  \item Reduction formulas:

\begin{equation}
G_{p,q}^{m,n}\left( z\left\vert 
\begin{array}{c}
a_{1},\ldots ,a_{p} \\ 
b_{1},\ldots ,b_{q-1},a_{1}%
\end{array}%
\right. \right) =G_{p-1,q-1}^{m,n-1}\left( z\left\vert 
\begin{array}{c}
a_{2},\ldots ,a_{p} \\ 
b_{1},\ldots ,b_{q-1}%
\end{array}%
\right. \right)\,,  
\end{equation}%
and%
\begin{equation}
G_{p,q}^{m,n}\left( z\left\vert 
\begin{array}{c}
a_{1},\ldots ,a_{p-1},b_{1} \\ 
b_{1},\ldots ,b_{q}%
\end{array}%
\right. \right) =G_{p-1,q-1}^{m-1,n}\left( z\left\vert 
\begin{array}{c}
a_{1},\ldots ,a_{p-1} \\ 
b_{2},\ldots ,b_{q}%
\end{array}%
\right. \right)\,.
\end{equation}

\item Derivative formulas:

\begin{eqnarray}
&&\frac{d}{dz}\left[ z^{1-a_{1}}\ G_{p,q}^{m,n}\left( z\left\vert 
\begin{array}{c}
a_{1},\ldots ,a_{p} \\ 
b_{1},\ldots ,b_{q}%
\end{array}%
\right. \right) \right] \\
&=&z^{-a_{1}}\ G_{p,q}^{m,n}\left( z\left\vert 
\begin{array}{c}
a_{1}-1,\ldots ,a_{p} \\ 
b_{1},\ldots ,b_{q}%
\end{array}%
\right. \right) ,\quad n\geq 1,  \notag
\end{eqnarray}%
and%
\begin{eqnarray}
&&\frac{d}{dz}\left[ z^{1-a_{p}}\ G_{p,q}^{m,n}\left( z\left\vert 
\begin{array}{c}
a_{1},\ldots ,a_{p} \\ 
b_{1},\ldots ,b_{q}%
\end{array}%
\right. \right) \right]  \\
&=&-z^{-a_{p}}\ G_{p,q}^{m,n}\left( z\left\vert 
\begin{array}{c}
a_{1},\ldots ,a_{p}-1 \\ 
b_{1},\ldots ,b_{q}%
\end{array}%
\right. \right) ,\quad n\leq p-1.  \notag
\end{eqnarray}

\item Translation formulas in the parameters:

\begin{equation}
z^{\alpha }G_{p,q}^{m,n}\left( z\left\vert 
\begin{array}{c}
a_{1},\ldots ,a_{p} \\ 
b_{1},\ldots ,b_{q}%
\end{array}%
\right. \right) =G_{p,q}^{m,n}\left( z\left\vert 
\begin{array}{c}
a_{1}+\alpha ,\ldots ,a_{p}+\alpha \\ 
b_{1}+\alpha ,\ldots ,b_{q}+\alpha%
\end{array}%
\right. \right)\,. 
\end{equation}

\item Relation with the generalized hypergeometric function $_{p}F_{q}$:

\begin{eqnarray}
&&_{p}F_{q}\left( \left. 
\begin{array}{c}
a_{1},\ldots ,a_{p} \\ 
b_{1},\ldots ,b_{q}%
\end{array}%
\right\vert -x\right)  \label{Meijer->p_F_q} \notag\\
&=&\frac{\prod_{\ell =1}^{q}\Gamma \left( b_{\ell }\right) }{\prod_{\ell
=1}^{p}\Gamma \left( a_{\ell }\right) }\, G_{p,q+1}^{1,p}\left( x\left\vert 
\begin{array}{c}
1-a_{1},\ldots ,1-a_{p} \\ 
0,1-b_{1},\ldots ,1-b_{q}%
\end{array}%
\right. \right)\,. \nonumber\\
\end{eqnarray}

\end{itemize}

\section{High PN accuracy spectral densities of energy and momentum losses at $O(G^3)$ in the rest frame of particle A}
\label{appD}

As a side project of our current works on gravitational wave emission  from scattering motions (notably  
\cite{Bini:2024tft}) we give in this appendix the $O(p_\infty^{30})$-accurate values of the spectral densities of energy and momentum losses at $O(G^3)$, computed in the rest frame of particle A.
Let
\beq
h_{ij}^{\rm A\, TT}=\frac{f_{ij}(u_{\rm ret}^{\rm A},\theta^{\rm A},\phi^{\rm A})}{R_{\rm A}}+O\left(\frac{1}{R_{\rm A}^2}\right)\,.
\eeq
denote the bremsstrahlung waveform recorded in the  rest frame of particle A (expressed  in terms of the retarded time and emission angle associated with the A rest frame, \cite{Bini:2024tft}).
We consider the A-frame Fourier transform of $f_{ij}^{\rm A\, TT}$
\beq
\hat f_{ij}(\omega^{\rm A},\theta^{\rm A},\phi^{\rm A})=\int_{-\infty}^{+\infty}  {d u_{\rm ret}^{\rm A}}e^{i\omega^{\rm A} u_{\rm ret}^{\rm A}}f_{ij}(u_{\rm ret}^{\rm A},\theta^{\rm A},\phi^{\rm A})\,.
\eeq
It is then convenient to work with a dimensionless rescaled version of the frequency $\omega_A$
that we shall denote as $u_A$ (it should not  be confused with the retarded time $u_{\rm ret}^{\rm A}$).
We also define the dimensionless time variable $T_A$  associated with $u_A$, namely 
\beq
\label{TA_def}
u_A\equiv\frac{\omega_A b}{p_\infty}= \frac{\omega_A b}{\g v}\,,\qquad T_A\equiv\frac{\gamma v}{b} u_{\rm ret}^{\rm A}\,.
\eeq
We then get 
\bea
\hat  f_{ij}(\om_{\rm A},\theta^{\rm A},\phi^{\rm A})&=&\frac{b}{\gamma v}\int d T_{\rm A} e^{i\omega^{\rm A} \frac{b}{\gamma v} T_{\rm A}}f_{ij}(T_{\rm A},\theta^{\rm A},\phi^{\rm A})\nonumber\\
&=&\frac{b}{\gamma v}\int d T_{\rm A} e^{iu_{\rm A}  T_{\rm A}}f_{ij}(T_{\rm A},\theta^{\rm A},\phi^{\rm A})\,.\nonumber\\
\eea
Hereafter, we often simplify the notation by deleting the A labels, i.e. by writing
\beq
u_{\rm A}=u\,,\qquad \Omega_{\rm A}=(\theta_{\rm A},\phi_{\rm A})=\Omega\,.
\eeq

Defining 
\beq
{\mathfrak F}(\om_{\rm A},\Omega^{\rm A})=(- i \om_A \hat  f_{ij}(\om_{\rm A},\Omega^{\rm A})) (- i \om_A \hat  f_{ij}(\om_{\rm A},\Omega^{\rm A}))^*
\eeq
the spectral density of radiated A-frame energy reads  
\bea
\frac{dE_A}{d\om_A}(\om_A)&=&  
\frac{1}{2\pi}
\frac{1}{32\pi G}\int d\Omega_{\rm A} {\mathfrak F}(\om_{\rm A},\Omega^{\rm A})\,,
\eea
while the spectral density of radiated A-frame momentum reads
\bea
\frac{d {\bf P}_A }{d\om_A}(\om_A)&=&   
\frac{1}{2\pi}\frac{1}{32\pi G}\int d\Omega_{\rm A}  {\mathfrak F}(\om_{\rm A},\Omega^{\rm A})\, {\bf n}_A\,.
\eea
Here we define the A-frame polar angles $(\theta_A, \phi_A)$ such that the direction of gravitational wave emission reads
\beq 
{\bf n}_A(\theta_A,\phi_A) = \cos\theta_A \,e_x^A +\sin\theta_A \cos\phi_A \, e^A_y +\sin\theta_A\sin\phi_A \, e_z^A  \,,  
\eeq
where (following \cite{kov-tho4}) $e_x^A$ is the direction of motion of body B in the A-frame, $e_y^A$ is in the direction of the impact parameter, while $e^A_z$ is orthogonal to the plane of motion. 

We found that the real integrand ${\mathfrak F}(\om_{\rm A},\Omega^{\rm A})$ depends on the angle $\phi_A$ only as $\sim c_0 +c_2 \cos(2\phi_A)+c_4 \cos(4\phi_A)+\ldots$. This implies that, after integrating over angles,  the   linear  momentum  spectral density $\frac{d {\bf P}_A }{d\om_A}(\om_A)$ is entirely directed 
along the axis $e_x^A$.

The corresponding integrated losses are
\be \label{EArad}
E_A^{\rm rad}=\int_{-\infty}^{+\infty} d\om_A\frac{dE_A}{d\om_A}(\om_A)\,,
\ee
and
\be \label{PArad}
P_A^{x \rm rad}=\int_{-\infty}^{+\infty} d\om_A\frac{dP_A^x }{d\om_A}(\om_A)\,.
\ee
Starting from the time-domain expression of the A-frame waveform (first computed in Ref. \cite{kov-tho4})
we computed the Fourier transform $\hat  f_{ij}(\om_{\rm A},\theta^{\rm A},\phi^{\rm A})$ and then
the angle-integrated spectral energy loss $\frac{dE_A}{d\om_A}(\om_A)$ to the thirtieth order
in the relative velocity $v= \pinf/\g$. It is convenient to express the result in terms of a
rescaled flux ${\mathcal F}(u)$ defined so that the {\it one-sided} integrand,
 $2d\om_A\frac{dE_A}{d\om_A}(\om_A)$, 
 of the total energy loss \eq{EArad} reads
\be \label{Fdef}
2d\om_A\frac{dE_A}{d\om_A}(\om_A)\bigg|_{\omega_A=\frac{\gamma v}{b}u} =   \frac{\gamma v}{b} \frac{G^3m_1^2 m_2^2}{b^3} du \, {\mathcal F}(u)\,,
\ee
Here, the extra factor 2 on the lhs is added because one must integrate both sides only on the positive real axis.

The velocity expansion of the dimensionless spectral density ${\mathcal F}(u)$ contains
only even powers of $v$, 
\be
{\mathcal F}(u)=\sum_{k=0}^{15}{\mathcal F}_{2k}(u)v^{2k}\,,
\ee
with expansion coefficients which are bilinear in $K_0(u)$ and $K_1(u)$, namely
\bea
{\mathcal F}_{2k}(u)&=&A_{00}^{(2k)}(u)K_0^2(u)+A_{01}^{(2k)}(u)K_0(u)K_1(u)\nonumber\\
&+&
A_{11}^{(2k)}(u)K_1^2(u)\,,
\eea
where  $A_{ij}^{(k)}(u)$ are  polynomials in $u$ whose orders increase with $k$. 
More precisely, $A_{00}^{(0)}(u),A_{11}^{(0)}(u)=P_4(u)$, $A_{01}^{(0)}(u)=P_3(u)$; $A_{00}^{(2)}(u),A_{11}^{(2)}(u)=P_6(u)$, $A_{01}^{(2)}(u)=P_5(u)$, etc. The values of these polynomials are given 
 in Tables \ref{tab:table1}, \ref{tab:table1} and \ref{tab:table1c} below.
They are also given in the ancillary file to this work.
We recall that the argument $u=u_A$ used here is defined as  $\frac{\omega_A b}{\g v}$.

As a check on our results, we have computed the total A-frame radiated energy and found it to agree
with the exact value of $E_A^{\rm rad}= \frac{G^3m_A^2m_B^2}{b^3}{\mathcal E}$ 
(first computed in \cite{Herrmann:2021lqe,Herrmann:2021tct}), namely
\begin{widetext}
\bea
{\mathcal E} &=& \int_0^{+\infty} du {\mathcal F}(u) \nonumber \\ 
&=&  \frac{37 \pi  v}{15}+\frac{2393 \pi  v^3}{840}+\frac{61703 \pi  v^5}{10080}+\frac{3131839 \pi  v^7}{354816}\nonumber\\
&+&\frac{513183289 \pi  v^9}{46126080}+\frac{60697345 \pi 
   v^{11}}{4612608}+\frac{588430385 \pi  v^{13}}{39207168}\nonumber\\
&+& \frac{12755740946147 \pi  v^{15}}{762814660608}+\frac{27966105533111 \pi  v^{17}}{1525629321216}+O\left(v^{18}\right)\,,
\eea
where we have shown only the first few terms of the exact result (the first fifteen terms where previously
computed by us in \cite{Bini:2021gat}). 

Finally, the only nonvanishing component $d\om_A\frac{dP_A^x }{d\om_A}(\om_A)$ of the spectral momentum loss is found not to carry any new information with respect to the energy spectrum. Indeed, we found that it is proportional to it and given by (when normalized as in \eq{Fdef})
\beq
{\mathcal F}_P^x(u)=\sqrt{\frac{\gamma-1}{\gamma+1}}{\mathcal F}(u)\,.
\eeq
The corresponding  integrated  radiated momentum agrees with, e.g., Eq. (5.13)  of Ref. \cite{Bini:2021gat}.

\begin{table}  
\caption{\label{tab:table1} Coefficients of the  PN-expansion of the A-frame  radiated energy spectrum at 3PM, $\frac{dE_A^{\rm rad}}{du}=\frac{1}{\pi} \frac{G^3 m_1^2m_2^2}{b^2} \frac{\gamma v}{b} 
\sum_{k=0}^{15}{\mathcal F}_{2k}(u)v^{2k}$, with ${\mathcal F}_{2k}(u)=A_{00}^{(2k)}(u)K_0^2(u)+A_{01}^{(2k)}(u)K_0(u)K_1(u)+A_{11}^{(2k)}(u)K_1^2(u)$. }
\begin{ruledtabular}
\begin{tabular}{l|ll}
    & $A_{00}^{(0)}$ &  $\frac{32}{5}\left(\frac{u^2}{3}+u^4\right)$\\
$0$ & $A_{01}^{(0)}$ &  $\frac{96}{5} u^3 $ \\ 
   & $A_{11}^{(0)}$   & $\frac{32}{5} (u^2+u^4)$\\
\hline
    & $A_{00}^{(2)}$ &  $\frac{16 u^6}{21}+\frac{16 u^4}{105}-\frac{160 u^2}{21} $\\
$2$ & $A_{01}^{(2)}$ &  $\frac{288 u^3}{35}-\frac{736 u^5}{105} $ \\ 
   & $A_{11}^{(2)}$   & $ \frac{16 u^6}{21}-\frac{104 u^4}{35}+\frac{160 u^2}{7} $\\
\hline
    & $A_{00}^{(4)}$ &  $-\frac{992}{315} u^{2}-\frac{592}{105} u^{4}+\frac{9752}{2835} u^{6}+\frac{32}{567} u^{8} $\\
$4$ & $A_{01}^{(4)}$ &  $ \frac{2144}{315} u^{3}-\frac{25792}{2835} u^{5}-\frac{32}{27} u^{7}$ \\ 
   & $A_{11}^{(4)}$   & $\frac{352}{9} u^{2}+\frac{22808}{2835} u^{4}+\frac{8152}{2835} u^{6}+\frac{32}{567} u^{8} $\\
\hline
    & $A_{00}^{(6)}$ &  $\frac{23648}{3465} u^{2}-\frac{35824}{3465} u^{4}+\frac{280094}{31185} u^{6}+\frac{15857}{31185} u^{8}+\frac{4}{1485} u^{10} $\\
$6$ & $A_{01}^{(6)}$ &  $ -\frac{3616}{385} u^{3}-\frac{569656}{31185} u^{5}-\frac{160163}{31185} u^{7}-\frac{208}{2079} u^{9}$ \\ 
   & $A_{11}^{(6)}$   & $\frac{19168}{385} u^{2}+\frac{516416}{31185} u^{4}+\frac{206834}{31185} u^{6}+\frac{14339}{31185} u^{8}+\frac{4}{1485} u^{10} $\\
\hline
    & $A_{00}^{(8)}$ &  $\frac{885344}{45045} u^{2}-\frac{2395024}{225225} u^{4}+\frac{3230374}{160875} u^{6}+\frac{7295669}{3378375} u^{8}+\frac{132532}{3378375} u^{10}+\frac{128}{1447875} u^{12} $\\
$8$ & $A_{01}^{(8)}$ &  $ -\frac{2715296}{75075} u^{3}-\frac{13389016}{375375} u^{5}-\frac{50489827}{3378375} u^{7}-\frac{764656}{1126125} u^{9}-\frac{1472}{289575} u^{11}$ \\ 
   & $A_{11}^{(8)}$   & $\frac{236576}{4095} u^{2}+\frac{4481456}{160875} u^{4}+\frac{9462994}{675675} u^{6}+\frac{95519}{51975} u^{8}+\frac{33844}{921375} u^{10}+\frac{128}{1447875} u^{12} $\\
\hline
    & $A_{00}^{(10)}$ &  $\frac{1541216}{45045} u^{2}-\frac{32624}{6825} u^{4}+\frac{31139132}{779625} u^{6}+\frac{67172618}{10135125} u^{8}+\frac{6949772}{30405375} u^{10}+\frac{1264}{675675} u^{12}+\frac{64}{30405375} u^{14} $\\
$10$ & $A_{01}^{(10)}$ &  $-\frac{15965344}{225225} u^{3}-\frac{651864016}{10135125} u^{5}-\frac{23864242}{675675} u^{7}-\frac{11806268}{4343625} u^{9}-\frac{209312}{4343625} u^{11}-\frac{5248}{30405375} u^{13}
 $ \\ 
   & $A_{11}^{(10)}$   & $ \frac{222304}{3465} u^{2}+\frac{39704584}{921375} u^{4}+\frac{38610844}{1447875} u^{6}+\frac{4944236}{921375} u^{8}+\frac{249748}{1216215} u^{10}+\frac{464}{259875} u^{12}+\frac{64}{30405375} u^{14}$\\
\end{tabular}
\end{ruledtabular}
\end{table}

\begin{table}  
\caption{\label{tab:table1b} See \ref{tab:table1}.
}
\begin{ruledtabular}
\begin{tabular}{l|ll}
    & $A_{00}^{(12)}$ &  $ \frac{38234848}{765765} u^{2}+\frac{664179056}{80405325} u^{4}+\frac{609846477308}{8442559125} u^{6}+\frac{3586085674}{216475875} u^{8}+\frac{13409764}{15049125} u^{10}+\frac{1070874916}{75983032125} u^{12}+\frac{4604272}{75983032125} u^{14}$\\
&& $+\frac{64}{1688511825} u^{16}$\\
$12$ & $A_{01}^{(12)}$ &  $ -\frac{691906016}{6185025} u^{3}-\frac{904026194128}{8442559125} u^{5}-\frac{15650879038}{216475875} u^{7}-\frac{5320118264}{649427625} u^{9}-\frac{18801541964}{75983032125} u^{11}-\frac{32693504}{15196606425} u^{13}-\frac{5056}{1206079875} u^{15}$ \\ 
   & $A_{11}^{(12)}$   & $\frac{53204192}{765765} u^{2}+\frac{527378010872}{8442559125} u^{4}+\frac{390447480532}{8442559125} u^{6}+\frac{108397714028}{8442559125} u^{8}+\frac{3457791644}{4469590125} u^{10}+\frac{991326004}{75983032125} u^{12}+\frac{4446448}{75983032125} u^{14}$\\
&& $+\frac{64}{1688511825} u^{16}
 $\\
\hline
    & $A_{00}^{(14)}$ &  $\frac{967086496}{14549535} u^{2}+\frac{44553349744}{1527701175} u^{4}+\frac{497489574458}{4113041625} u^{6}+\frac{5743861651901}{160408623375} u^{8}+\frac{52138451363}{19249034805} u^{10}+\frac{3173164499}{47140493400} u^{12}+\frac{55198609}{97053957000} u^{14}$\\
&&$+\frac{455251}{320817246750} u^{16}+\frac{2}{3749811975} u^{18}
 $\\
$14$ & $A_{01}^{(14)}$ &  $ -\frac{241339616992}{1527701175} u^{3}-\frac{2966753846488}{17823180375} u^{5}-\frac{109459029181}{822608325} u^{7}-\frac{9833194123268}{481225870125} u^{9}-\frac{1055131768589}{1154942088300} u^{11}-\frac{1547366591}{113229616500} u^{13}$\\
&&$-\frac{63589199}{962451740250} u^{15}-\frac{3152}{41247931725} u^{17}
$ \\ 
   & $A_{11}^{(14)}$   & $\frac{63368224}{855855} u^{2}+\frac{10005565904}{116491375} u^{4}+\frac{1324684450874}{17823180375} u^{6}+\frac{95116552793}{3564636075} u^{8}+\frac{193685786059}{84922212375} u^{10}+\frac{1403038676257}{23098841766000} u^{12}+\frac{1238993491}{2309884176600} u^{14}$\\
&&$+\frac{362519}{262486838250} u^{16}+\frac{2}{3749811975} u^{18}
 $\\
\hline
    & $A_{00}^{(16)}$ &  $ \frac{1216661536}{14549535} u^{2}+\frac{267653884592}{4583103525} u^{4}+\frac{54952443734834}{288735522075} u^{6}+\frac{20058843936491}{288735522075} u^{8}+\frac{4280998872419}{618718975875} u^{10}+\frac{25253447517269}{103944787947000} u^{12}$\\
&&$+\frac{337766531953}{103944787947000} u^{14}+\frac{1260100943}{77958590960250} u^{16}+\frac{2937982}{116937886440375} u^{18}+\frac{64}{10630716949125} u^{20}
$\\
$16$ & $A_{01}^{(16)}$ &  $-\frac{955109600096}{4583103525} u^{3}-\frac{5432756646472}{22210424775} u^{5}-\frac{7243357382909}{32081724675} u^{7}-\frac{3929659712072}{88388425125} u^{9}-\frac{140892100269227}{51972393973500} u^{11}-\frac{149172517973}{2474875903500} u^{13}$\\
&&$-\frac{30142913}{59555837250} u^{15}-\frac{13319576}{8995222033875} u^{17}-\frac{256}{236238154425} u^{19}
 $ \\ 
   & $A_{11}^{(16)}$   & $\frac{59755616}{765765} u^{2}+\frac{32681704037984}{288735522075} u^{4}+\frac{244034023054}{2170943775} u^{6}+\frac{72104443743799}{1443677610375} u^{8}+\frac{10514507163607}{1856156927625} u^{10}+\frac{8908446852703}{41577915178800} u^{12}$\\
&&$+\frac{12489841249}{4157791517880} u^{14}+\frac{3609992339}{233875772880750} u^{16}+\frac{2874974}{116937886440375} u^{18}+\frac{64}{10630716949125} u^{20}
 $\\
\hline
    & $A_{00}^{(18)}$ &  $\frac{4840422944}{47805615} u^{2}+\frac{10152535152304}{105411381075} u^{4}+\frac{9454139678145704}{33204585038625} u^{6}+\frac{152535831520652}{1229799445875} u^{8}+\frac{81453937212064}{5242829216625} u^{10}+\frac{2799488660851}{3881055394125} u^{12}$\\
&&$+\frac{163614684321623}{11953650613905000} u^{14}+\frac{374166171979}{3448168446318750} u^{16}+\frac{22895301998}{67239284703215625} u^{18}+\frac{23391824}{67239284703215625} u^{20}+\frac{32}{574694741053125} u^{22}
 $\\
$18$ & $A_{01}^{(18)}$ &  $-\frac{27673801976032}{105411381075} u^{3}-\frac{11402751015271744}{33204585038625} u^{5}-\frac{11935974246697772}{33204585038625} u^{7}-\frac{8691906334862104}{99613755115875} u^{9}-\frac{2053081414878682}{298841265347625} u^{11}$\\
&&$-\frac{12737341722449}{60988013336250} u^{13}-\frac{117364295407867}{44826189802143750} u^{15}-\frac{128906265112}{9605612100459375} u^{17}-\frac{1699134176}{67239284703215625} u^{19}-\frac{2624}{213458046676875} u^{21}
 $ \\ 
   & $A_{11}^{(18)}$   & $ \frac{27301225312}{334639305} u^{2}+\frac{4786820572883816}{33204585038625} u^{4}+\frac{21948751897528}{135528918525} u^{6}+\frac{572935348006816}{6640917007725} u^{8}+\frac{148132470366512}{11953650613905} u^{10}+\frac{37225425774913}{59768253069525} u^{12}$\\
&&$+\frac{742786600096999}{59768253069525000} u^{14}+\frac{391765471967}{3842244840183750} u^{16}+\frac{4411424734}{13447856940643125} u^{18}+\frac{22980416}{67239284703215625} u^{20}+\frac{32}{574694741053125} u^{22}
$\\
\hline
    & $A_{00}^{(20)}$ &  $ \frac{199549245664}{1673196525} u^{2}+\frac{9764857623088}{68207364225} u^{4}+\frac{1641965775544509392}{4017754789673625} u^{6}+\frac{49012919505720124}{236338517039625} u^{8}+\frac{1903731817581247496}{60266321845104375} u^{10}$\\
&&$+\frac{334904293968193186}{180798965535313125} u^{12}+\frac{25930496940068909}{556304509339425000} u^{14}+\frac{17132205851197}{32556836530968750} u^{16}+\frac{107724646980226}{40679767245445453125} u^{18}$\\
&&$+\frac{224663483288}{40679767245445453125} u^{20}+\frac{1345072}{347690318337140625} u^{22}+\frac{64}{149010136430203125} u^{24}
$\\
$20$ & $A_{01}^{(20)}$ &  $ -\frac{370945963204832}{1159525191825} u^{3}-\frac{1866351240994389472}{4017754789673625} u^{5}-\frac{34979088636028}{64309800555} u^{7}-\frac{9524631180898201616}{60266321845104375} u^{9}-\frac{2793320184131759948}{180798965535313125} u^{11}$\\
&&$-\frac{40697934644030207}{66962579827893750} u^{13}-\frac{283210361675487461}{27119844830296968750} u^{15}-\frac{3264882343848404}{40679767245445453125} u^{17}-\frac{3618884910856}{13559922415148484375} u^{19}$\\
&&$-\frac{510977344}{1506658046127609375} u^{21}-\frac{10816}{94824632273765625} u^{23}
$ \\ 
   & $A_{11}^{(20)}$   & $\frac{6167532064}{72747675} u^{2}+\frac{717617592078933608}{4017754789673625} u^{4}+\frac{901260951613981888}{4017754789673625} u^{6}+\frac{2810388801812460928}{20088773948368125} u^{8}+\frac{48857077588558216}{1986801819069375} u^{10}$\\
&&$+\frac{567056362248643441}{361597931070626250} u^{12}+\frac{1505165417688550327}{36159793107062625000} u^{14}+\frac{2832028348273429}{5811395320777921875} u^{16}+\frac{102403591879046}{40679767245445453125} u^{18}$\\
&&$+\frac{16757096072}{3129212865034265625} u^{20}+\frac{3975952}{1043070955011421875} u^{22}+\frac{64}{149010136430203125} u^{24}
 $\\
\end{tabular}
\end{ruledtabular}
\end{table}

\begin{table}  
\caption{\label{tab:table1c} See \ref{tab:table1}.
}
\begin{ruledtabular}
\begin{tabular}{l|ll}
\hline
    & $A_{00}^{(22)}$ &  $ \frac{138121194272}{1003917915} u^{2}+\frac{230926708047248}{1159525191825} u^{4}+\frac{6831720750861217138}{12053264369020875} u^{6}+\frac{3964026888009660151}{12053264369020875} u^{8}+\frac{10734813118303152454}{180798965535313125} u^{10}$\\
&&$+\frac{1845038794817382917}{433917517284751500} u^{12}+\frac{36287858835600971}{267850319311575000} u^{14}+\frac{110285140133743067}{54239689660593937500} u^{16}+\frac{19034122266002267}{1301752551854254500000} u^{18}$\\
&&$+\frac{12004414078169}{244078603472672718750} u^{20}+\frac{2889194543}{40679767245445453125} u^{22}+\frac{38765938}{1098353715627027234375} u^{24}+\frac{8}{2852866793836434375} u^{26}
$\\
$22$ & $A_{01}^{(22)}$ &  $ -\frac{23201056084832}{61027641675} u^{3}-\frac{7345590336598137608}{12053264369020875} u^{5}-\frac{634470727221493499}{803550957934725} u^{7}-\frac{48543744264598270504}{180798965535313125} u^{9}-\frac{34326059254774142599}{1084793793211878750} u^{11}$\\
&&$-\frac{73690227510294553}{47578675140871875} u^{13}-\frac{267301770040765541}{7748527094370562500} u^{15}-\frac{5388608317539481}{14792642634707437500} u^{17}-\frac{13489885200283}{7396321317353718750} u^{19}$\\
&&$-\frac{29596144306}{7178782455078609375} u^{21}-\frac{806517454}{219670743125405446875} u^{23}-\frac{4192}{4754777989727390625} u^{25}
$ \\ 
   & $A_{11}^{(22)}$   & $\frac{88029018016}{1003917915} u^{2}+\frac{2607781020372568672}{12053264369020875} u^{4}+\frac{3626025590836924766}{12053264369020875} u^{6}+\frac{1443756504976386533}{6696257982789375} u^{8}+\frac{3500727809438547682}{77485270943705625} u^{10}$\\
&&$+\frac{15333365220089683411}{4339175172847515000} u^{12}+\frac{87904585329932533}{737954961368625000} u^{14}+\frac{568236625586326}{305862911619890625} u^{16}+\frac{10423951630297}{759039388836300000} u^{18}$\\
&&$+\frac{1918269573533}{40679767245445453125} u^{20}+\frac{76010981173}{1098353715627027234375} u^{22}+\frac{38283302}{1098353715627027234375} u^{24}+\frac{8}{2852866793836434375} u^{26}
 $\\
\hline
    & $A_{00}^{(24)}$ &  $\frac{22731815702432}{145568097675} u^{2}+\frac{115603050920726128}{437140997318025} u^{4}+\frac{45117925910952854980534}{59073048672571308375} u^{6}+\frac{29498256429730204303753}{59073048672571308375} u^{8}$\\
&&$+\frac{92798171531991963534586}{886095730088569625625} u^{10}+\frac{5583706789992195513491}{625479338886049147500} u^{12}+\frac{3365701520757398403943}{9666498873693486825000} u^{14}+\frac{3574705737283178077}{539206326625904437500} u^{16}$\\
&&$+\frac{405279261932776373801}{6379889256637701304500000} u^{18}+\frac{52353104538342463}{170889890802795570656250} u^{20}+\frac{85801830424603}{119622923561956899459375} u^{22}+\frac{3990890959582}{5383031560288060475671875} u^{24}$\\
&&$+\frac{131576072}{489366505480732770515625} u^{26}+\frac{512}{32624433698715518034375} u^{28}
 $\\
$24$ & $A_{01}^{(24)}$ &  $-\frac{193661419297856992}{437140997318025} u^{3}-\frac{46043637666221020423544}{59073048672571308375} u^{5}-\frac{1453914081765252799547}{1312734414946029075} u^{7}-\frac{383647190582880696428896}{886095730088569625625} u^{9}$\\
&&$-\frac{319656138559373050266541}{5316574380531417753750} u^{11}-\frac{47257910400877740618083}{13291435951328544384375} u^{13}-\frac{26232831968971568571821}{265828719026570887687500} u^{15}-\frac{98095661041066510163}{72498741552701151187500} u^{17}$\\
&&$-\frac{5581010221842893798}{598114617809784497296875} u^{19}-\frac{19070642631431098}{598114617809784497296875} u^{21}-\frac{2300004655006}{45235559330151768703125} u^{23}$\\
&&$-\frac{2287755488}{69909500782961824359375} u^{25}-\frac{14464}{2509571822978116771875} u^{27}
 $ \\ 
   & $A_{11}^{(24)}$   & $ \frac{175359843808}{1940907969} u^{2}+\frac{15194805030800967293776}{59073048672571308375} u^{4}+\frac{23200489852455602449538}{59073048672571308375} u^{6}+\frac{10459835589057082288667}{32818360373650726875} u^{8}$\\
&&$+\frac{207338776930837470456826}{2658287190265708876875} u^{10}+\frac{6740868007239141331783}{924621631396768305000} u^{12}+\frac{160394890732671759827483}{531657438053141775375000} u^{14}+\frac{265009247773036462667}{44304786504428481281250} u^{16}$\\
&&$+\frac{1129254064381335724451}{19139667769913103913500000} u^{18}+\frac{173903961494652086}{598114617809784497296875} u^{20}+\frac{3726162538260727}{5383031560288060475671875} u^{22}+\frac{780704894114}{1076606312057612095134375} u^{24}$\\
&&$+\frac{130169672}{489366505480732770515625} u^{26}+\frac{512}{32624433698715518034375} u^{28}
$\\
\hline
    & $A_{00}^{(26)}$ &  $ \frac{789497272234592}{4512611027925} u^{2}+\frac{4596440091268254832}{13551370916858775} u^{4}+\frac{7568803759095722524636}{7536067937653129875} u^{6}+\frac{148776310059357091351718}{203473834316634506625} u^{8}$\\
&&$+\frac{48637507979688620220412}{277464319522683418125} u^{10}+\frac{6938788342638232540448}{398100980184719686875} u^{12}+\frac{891099779139418716613927}{1098758705309826335775000} u^{14}+\frac{13003594981011664262176}{686724190818641459859375} u^{16}$\\
&&$+\frac{159616612887594572396437}{692217984345190591538250000} u^{18}+\frac{599618301729406061621}{403793824201361178397312500} u^{20}+\frac{1514380360205710136}{302845368151020883797984375} u^{22}$\\
&&$+\frac{109817796696688}{13041188102197071551109375} u^{24}+\frac{1586655202096}{247782573941744359471078125} u^{26}+\frac{430053856}{247782573941744359471078125} u^{28}$\\
&&$+\frac{256}{3363564352150375920421875} u^{30}
$\\
$26$ & $A_{01}^{(26)}$ &  $-\frac{6886863104740471648}{13551370916858775} u^{3}-\frac{3483141476090777766256}{3569716391519903625} u^{5}-\frac{16164736259909734889558}{10709149174559710875} u^{7}-\frac{20614165481495780467348}{30829368835853713125} u^{9}$\\
&&$-\frac{4846005875774071839112}{45105037163785974375} u^{11}-\frac{108470408870665347241271}{14457351385655609681250} u^{13}-\frac{173017710864121696969661}{686724190818641459859375} u^{15}-\frac{372966212459476870224091}{86527248043148823942281250} u^{17}$\\
&&$-\frac{93666116994953020177391}{2422762945208167070383875000} u^{19}-\frac{1083110701838558012}{5938144473549429094078125} u^{21}-\frac{1202966117036772124}{2725608313359187954181859375} u^{23}$\\
&&$-\frac{6674225083336}{13041188102197071551109375} u^{25}-\frac{6724566656}{27531397104638262163453125} u^{27}-\frac{52736}{1633731256758754018490625} u^{29}
 $ \\ 
   & $A_{11}^{(26)}$   & $ \frac{14441703673952}{155607276825} u^{2}+\frac{20418835843447440844616}{67824611438878168875} u^{4}+\frac{101989622562156844731004}{203473834316634506625} u^{6}+\frac{106840478952909326209996}{234777501134578276875} u^{8}$\\
&&$+\frac{1171380087782820390448852}{9156322544248552798125} u^{10}+\frac{45244603486308867471913}{3231643250911253928750} u^{12}+\frac{3807354042065751449217887}{5493793526549131678875000} u^{14}+\frac{617403067392999225892}{36571110753655462359375} u^{16}$\\
&&$+\frac{1026932933782535793222439}{4845525890416334140767750000} u^{18}+\frac{140934419172894746038}{100948456050340294599328125} u^{20}+\frac{13038905593086765338}{2725608313359187954181859375} u^{22}$\\
&&$+\frac{31136983799756}{3812039599103759376478125} u^{24}+\frac{518868897952}{82594191313914786490359375} u^{26}+\frac{1278192416}{743347721825233078413234375} u^{28}$\\
&&$+\frac{256}{3363564352150375920421875} u^{30}
$\\
\hline
    & $A_{00}^{(28)}$ &  $\frac{875145464391392}{4512611027925} u^{2}+\frac{28692884397587869328}{67756854584293875} u^{4}+\frac{1691745290672435991120668}{1308046077749793256875} u^{6}+\frac{9506814253980196831066946}{9156322544248552798125} u^{8}$\\
&&$+\frac{7713687530535308372495164}{27468967632745658394375} u^{10}+\frac{528139608636806484446231}{16481380579647395036625} u^{12}+\frac{19345898893610005968287}{11098572780907336725000} u^{14}+\frac{1012016374898210408204}{20809823964201256359375} u^{16}$\\
&&$+\frac{85833877292121815579}{118025231772411013050000} u^{18}+\frac{43330305867468426723967}{7268288835624501211151625000} u^{20}+\frac{117150021962742087743}{4360973301374700726690975000} u^{22}$\\
&&$+\frac{232004047450754431}{3555141278294592983715468750} u^{24}+\frac{790577954597104}{9734315404854242693506640625} u^{26}+\frac{214402978984}{4610991507562536012713671875} u^{28}$\\
&&$+\frac{453495712}{47173990038909022283916796875} u^{30}+\frac{128}{396420084360580019192578125} u^{32}

 $\\
$28$ & $A_{01}^{(28)}$ &  $-\frac{38995906600648639904}{67756854584293875} u^{3}-\frac{1568942186739045088015888}{1308046077749793256875} u^{5}-\frac{211266875634689159579914}{105245086715500606875} u^{7}$\\
&&$-\frac{27353804323694142943330456}{27468967632745658394375} u^{9}-\frac{3006514681467466450934134}{16481380579647395036625} u^{11}-\frac{4057701077695542063869821}{274689676327456583943750} u^{13}-\frac{134275512321977171029519}{228908063606213819953125} u^{15}$\\
&&$-\frac{350616364055830627460663}{28842416014382941314093750} u^{17}-\frac{330027259330570258820161}{2422762945208167070383875000} u^{19}-\frac{912791336175170588417}{1090243325343675181672743750} u^{21}$\\
&&$-\frac{229257545681633477809}{81768249400775638625455781250} u^{23}-\frac{337355954747939048}{68140207833979698854546484375} u^{25}-\frac{138183098407624}{32276940552937752088995703125} u^{27}$\\
&&$-\frac{24349412096}{15724663346303007427972265625} u^{29}-\frac{12416}{79284016872116003838515625} u^{31}
 $ \\ 
   & $A_{11}^{(28)}$   & $\frac{13842435028768}{145568097675} u^{2}+\frac{454822710628184601462632}{1308046077749793256875} u^{4}+\frac{1149075091475118140208476}{1831264508849710559625} u^{6}+\frac{29628524697134942243156}{46955500226915655375} u^{8}$\\
&&$+\frac{16567062699346820224053628}{82406902898236975183125} u^{10}+\frac{417608713557307151056939}{16481380579647395036625} u^{12}+\frac{8075902722379716894597419}{5493793526549131678875000} u^{14}+\frac{29425088012532655789}{686381000318482218750} u^{16}$\\
&&$+\frac{3206891718513900931193881}{4845525890416334140767750000} u^{18}+\frac{17732972435528737759}{3192046041117479671125000} u^{20}+\frac{151596732490556699963}{5946781774601864627305875000} u^{22}$\\
&&$+\frac{103985251398356231}{1655227720663474466102343750} u^{24}+\frac{646767703118128}{8176824940077563862545578125} u^{26}+\frac{1869579635032}{40884124700387819312727890625} u^{28}$\\
&&$+\frac{34600736}{3628768464531463252608984375} u^{30}+\frac{128}{396420084360580019192578125} u^{32}
 $\\
\end{tabular}
\end{ruledtabular}
\end{table}

\end{widetext}

\end{document}